\documentclass[
aps,%
12pt,%
final,%
notitlepage,%
oneside,%
onecolumn,%
nobibnotes,%
nofootinbib,
superscriptaddress,%
noshowpacs,%
centertags]%
{revtex4}

\usepackage{graphicx}
\usepackage{color}

\newcommand{\pT}{\ensuremath{p_{\rm t}}} 
\newcommand{\mT}{\ensuremath{m_{\rm t}}} 
\newcommand{\sqsNN}{\sqrt{s_{_{NN}}}}
\begin{document}
\selectlanguage{english} 
\title{Physics with the ALICE experiment}
\author{\firstname{Yuri} \surname{Kharlov, for the ALICE collaboration}}
\affiliation{Institute for High Energy Physics, Protvino, 142281 Russia}
%
%
\begin{abstract}
  ALICE experiment at LHC collects data in pp collisions at
  $\sqrt{s}$=0.9, 2.76 and 7~TeV and in PbPb collisions at
  2.76~TeV. Highlights of the detector performance and an overview of
  experimental results measured with ALICE in pp and AA collisions
  are presented in this paper. Physics with proton-proton
  collisions is focused on hadron spectroscopy at low and
  moderate $\pT$. Measurements with lead-lead collisions are shown in
  comparison with those in pp collisions, and the properties of hot
  quark matter are discussed.
\end{abstract}
\maketitle
%

\section{Introduction}

ALICE is a dedicated experiment built to exploit the unique physics
potential of heavy-ion interactions at LHC energies
\cite{Aamodt:2008zz}. Properties of strongly interacting matter at
extreme energy density are explored via a comprehensive studies of
hadron, muon, electron and photon production in the collisions of
heavy nuclei and their comparison with proton-proton collisions.

Presently, the ALICE collaboration consists of about 1600 members from
33 countries. Russian nuclear-physics community takes an active part
in ALICE since the very beginning, now counting 134 members from 12
institutes. Russian institutes contribute in almost every major
sub-detectors of the ALICE experiment, and also take part in physics
analysis of data collected in 2010--2011.

The ALICE experiment has collected a rich sample of data with
proton-proton and lead-lead collisions. In 2010 and beginning of 2011,
about $10^9$~events with the minimum bias trigger were recorded,
corresponding to the integrated luminosity $\int{\cal L}dT =
16~\mbox{nb}^{-1}$. Rare-event triggers on muons, jets and photons
were dominant in data taking with the proton beams at collision energy
$\sqrt{s}=7$~TeV in the second half of 2011, with the delivered
integrated luminosity $\int{\cal L}dT = 4.9~\mbox{pb}^{-1}$. Limited
data samples with the proton beams at collision energies $\sqrt{s}=0.9$
and 2.76~TeV have been also recorded with integrated luminosities
$\int{\cal L}dT = 0.14\mbox{~and~}1.3~\mbox{nb}^{-1}$ respectively.

Among rare-event triggers used in data taking in 2011, one has to
mention the trigger on the MUON detector selecting events with muons
in the high-rapidity range to enrich statistics for $J/\psi$ and
$\Upsilon$ signals (this trigger was in operation since 2010). A
trigger based on the electromagnetic calorimeter (EMCAL) was selecting
events with high-energy photons and jets in the central
barrel. Another ALICE calorimeter, a photon spectrometer PHOS, has
provided a trigger on photons with a moderate energy threshold, to
enhance a data sample for neutral meson and direct photon studies.

The first run with lead-lead beams at collision energy
$\sqsNN=2.76$~TeV was taken with ALICE in November 2010. The
delivered integrated luminosity was $\int{\cal L}dT =
10~\mu\mbox{b}^{-1}$. The dominant trigger in 2010 was a minimum bias
one. In November 2011, the LHC has delivered 10 times more data, and
the ALICE experiment has restricted the minimum-bias trigger share in
favor of several rare-event triggers with the total life time
80\%. Detector VZERO has deployed triggers on the most central events
with selected centralities $0-10\%$ and semi-central events with
centralities $20-60\%$. A trigger on ultra-peripheral collisions was
realized on SPD and TOF detectors. Other triggers implemented earlier
in pp collisions on EMCAL, PHOS and MUON detectors, were also active
in the PbPb run 2011.

\section{Hadron production in proton-proton collisions}

Measurements of identified hadron spectra are considered as an
important test of various non-perturbative models of hadron production
at high energies, as well as those of perturbative QCD
calculations. ALICE performs extensive studies of hadron production
due to its powerful particle identification capabilities
\cite{Aamodt:2008zz}. Charged particles are identified by several
tracking detectors covering complimentary kinematic ranges. Barrel
tracking detectors are embedded into a solenoidal magnet with magnetic
field of 0.5~T. This is a relatively soft magnetic field which allows
to reconstruct charged tracks at transverse momenta starting from
$p_{\rm t}>50$~MeV/$c$.  Inner Tracking System (ITS) and Time
Projection Chamber (TPC) can identify charged particles in the full
$2\pi$ azimuthal angle and pseudorapidity range $|\eta|<0.9$, via
measurements of their ionization loss $dE/dx$. Time-of-flight
measurements, provided by the TOF detector in the same solid
angle as ITS and TPC, can discriminate charged pions, kaons and
protons in a higher momentum range. The limited-acceptance High-Momentum
Particle Identification detector (HMPID) is a Cherenkov detector
covering a solid angle $\Delta\phi=60^\circ$ and $|\eta|<0.6$ is used
to identify charged particles at a higher momentum range, up to
$p=5$~GeV/$c$. Transition Radiation Detector (TRD) is another
barrel detector surrounding TPC, which is designed to identify
electrons and at present covers about a half of the complete azimuthal
angle.

Photons and neutral mesons decaying into photons are detected and
identified by two electromagnetic calorimeters. A precise Photon
Spectrometer (PHOS) is a high-granularity calorimeter built of lead
tungstate crystals (PbWO$_4$). Its small Moli\'ere radius, high density
and high light yield allow to detect photons with the best possible
energy resolution in the energy range up to $E<100$~GeV in the
azimuthal angle range $\Delta\phi=60^\circ$ and $|\eta|<0.13$. Its high
spatial resolution provides measurements of neutral pions via
invariant mass spectrum at transverse momenta $0.6<p_{\rm
  t}<50$~GeV/$c$. Another, wide-aperture Electromagnetic Calorimeter
(EMCAL) is a sampling-type calorimeters built of lead-scintillator
modules. Its primary goal is to trigger jets and measure a neutral
component of jets. Dynamic range of EMCAL covers energies up to
250~GeV, and granularity of this calorimeter allows to reconstruct
$\pi^0$ mesons at transverse momenta $1<\pT<20$~GeV/$c$.

Muon identification is provided in ALICE by the muon arm which is
installed in the forward rapidity range $2.5<y<4$. This muon detector
is a magnet spectrometer consisting of a set of proportional chambers
in the dipole magnetic field. Hadronic background is suppressed by 
the hadron absorber installed in front of the muon spectrometer.

\begin{figure}[ht]
  \centering
  \includegraphics[width=0.5\hsize]{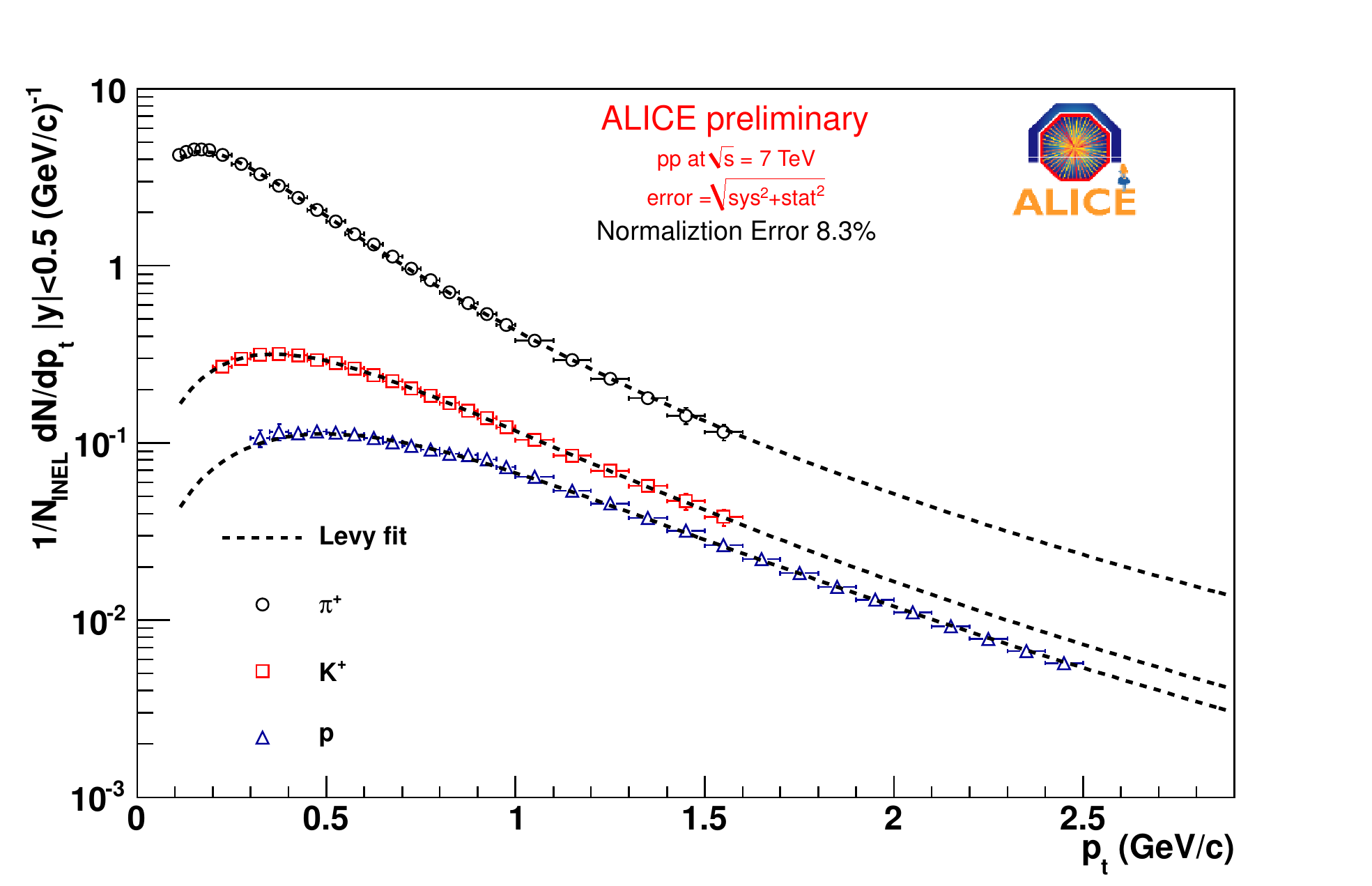}
  \caption{Transverse momentum spectra of $\pi^-$, $K^-$, $\bar{p}$ in
    pp collisions at $\sqrt{s} = 7$~TeV. The lines are the
    Levy-Tsallis fits.} 
  \label{fig:pp-piKp_spectra}
\end{figure}

\begin{figure}[ht]
  \includegraphics[width=0.48\hsize]{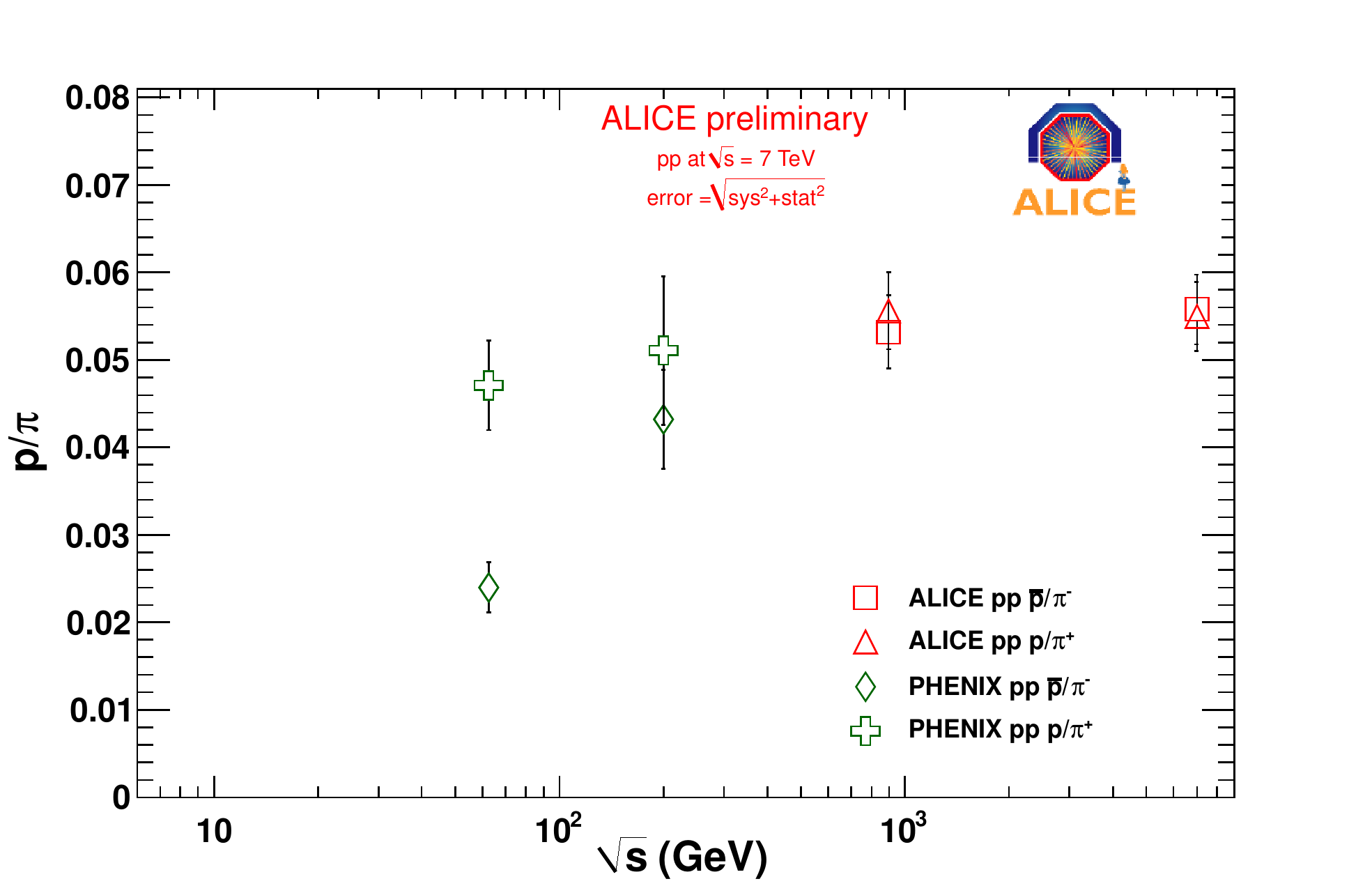}
  \hfil
  \includegraphics[width=0.48\hsize]{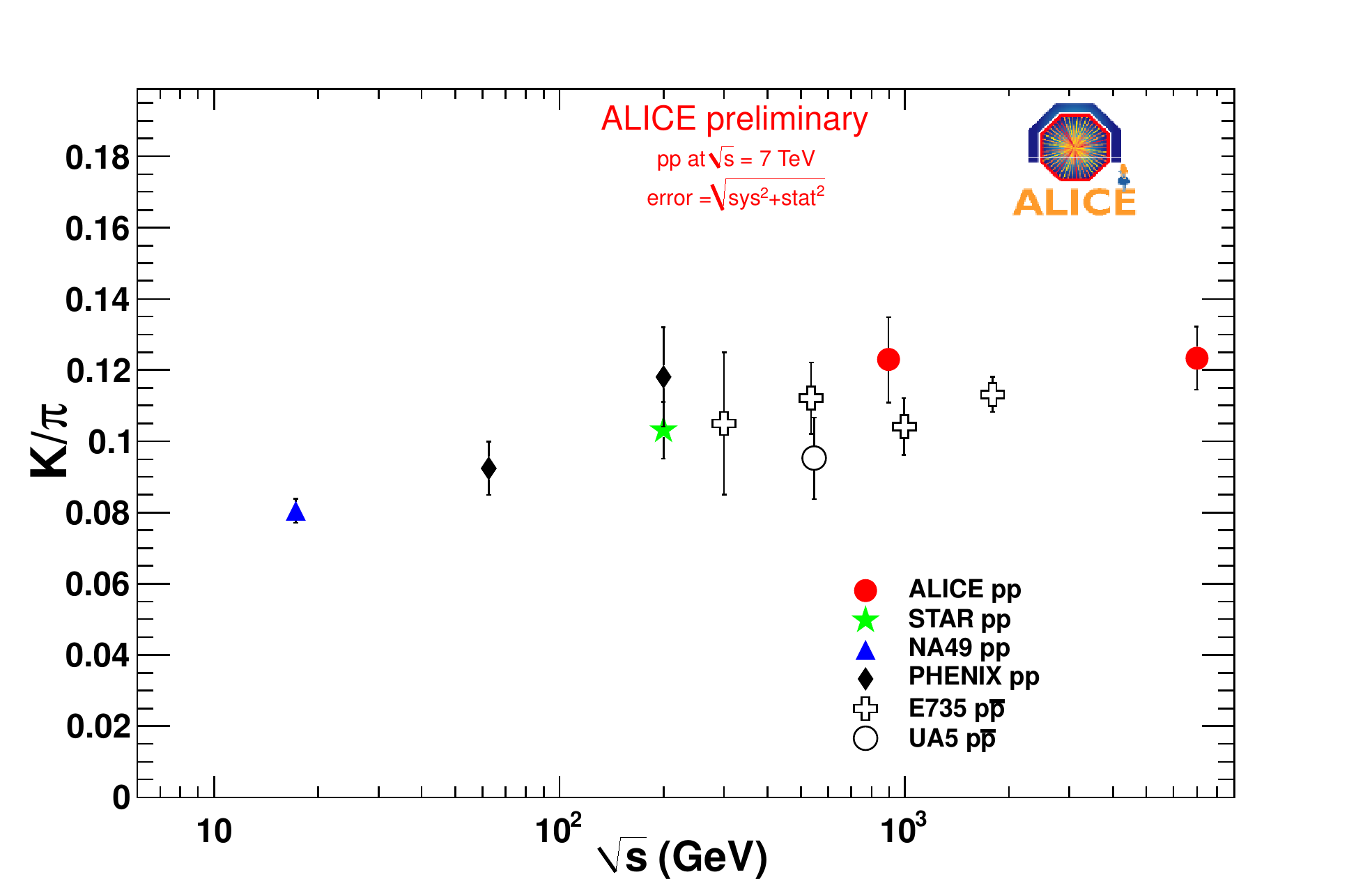}
  \caption{Integrated yield ratio of $K/\pi$ (left) and $\bar{p}/\pi^-$
    (right) as a function of collision energy.} 
  \label{fig:pp-k2pi-p2pi}
\end{figure}

Using charged hadron identification in ITS, TPC and TOF,
ALICE has measured production spectra $dN/d\pT$ of identified charged
hadron ($\pi^\pm$, $K^\pm$, $p$, $\bar{p})$ in the minimum bias pp
collisions at collision energies $\sqrt{s}=0.9$~\cite{PIDhadron900GeV}
and $7$~TeV~\cite{PIDhadron7TeV} (Fig.\ref{fig:pp-piKp_spectra}).
The spectra were fitted with the Tsallis function~\cite{Tsallis:1987eu}
\begin{equation}
   \displaystyle
\!\!\!\!\!\!\!\!\!\!\!\!\!\!\!
    E \frac{{\rm d}^3 \sigma}{{\rm d}p^3} = 
    \displaystyle
    \frac{\sigma_{pp}}{2\pi}\frac{{ \rm d}N}{{\rm d}y} 
    \frac{c \cdot (n-1)(n-2)}{nC\left[ nC+m(n-2)\right]}  
    \displaystyle\left(1+\frac{\mT-m}{nC}\right)^{-n},
    \label{eq:Tsallis}
\end{equation}
where the fit parameters are ${\rm d}N/{\rm d}y$, $C$ and $n$,
$\sigma_{\rm pp}$ is the proton-proton inelastic cross section, $m$ is
the meson rest mass and $\mT=\sqrt{m^2+\pT^2}$ is the transverse
mass. The integrated yield at $y=0$, defined by the Tsallis parameter
${\rm d}N/{\rm d}y$, was evaluated from the ALICE data, and thus the
total yields of charged pions, kaons and protons was found. The ratios
of integrated yields $K^\pm/\pi^\pm$, $\bar{p}/\pi^-$ and $p/\pi^+$ in
pp collisions at $\sqrt{s}=0.9$ and $7$~TeV were compared with those
measured at lower collision energies, as shown in
Fig.\ref{fig:pp-k2pi-p2pi}.
A trend of slight increase of $K^\pm/\pi^\pm$ ratio with $\sqrt{s}$
can be observed. ALICE data also suggest that baryon-antibaryon
asymmetry, observed at RHIC, vanishes at LHC energies, as expected.

\begin{figure}[ht]
  \centering
  \includegraphics[width=0.5\hsize]{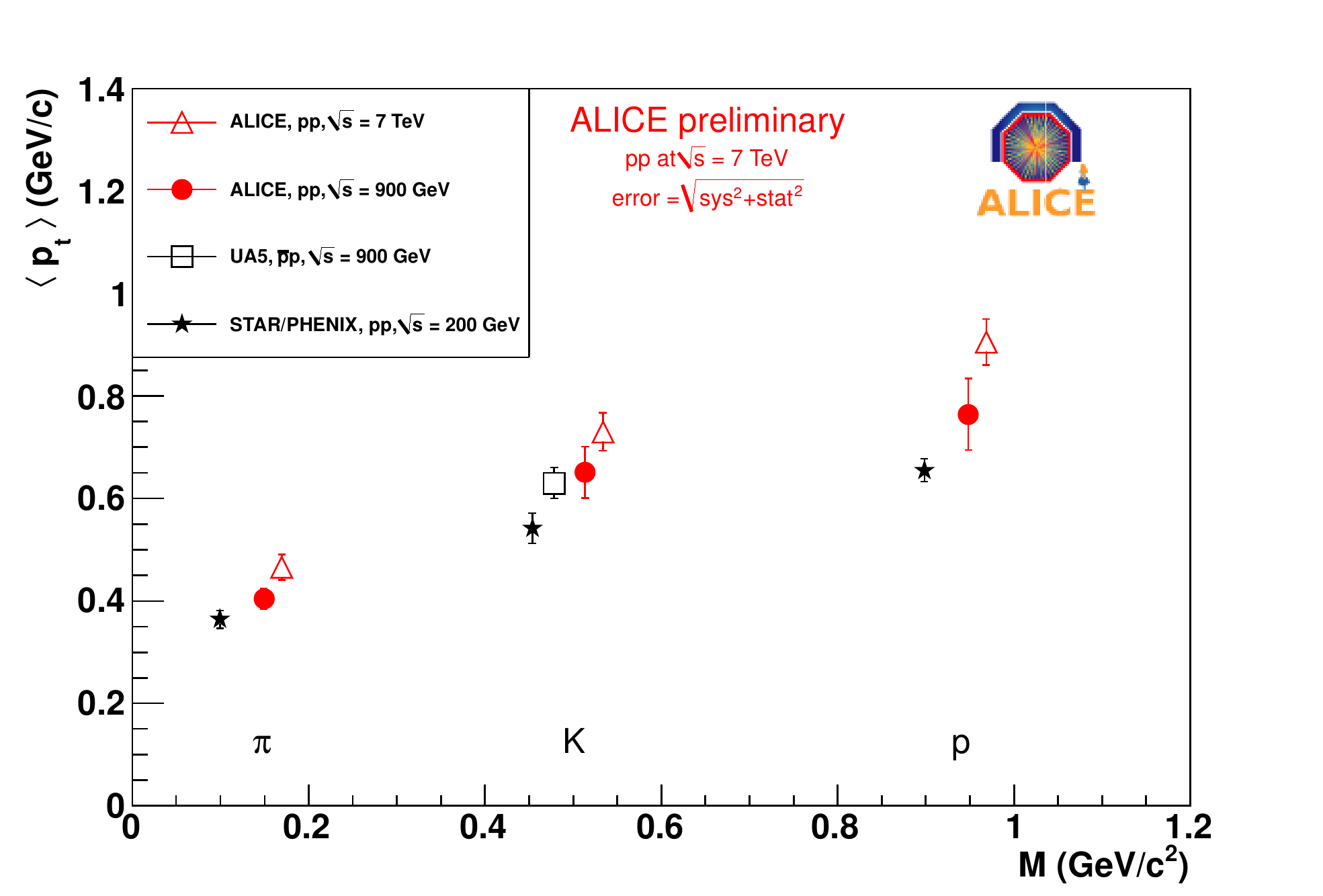}
  \caption{Mean $p_{\rm t}$ for charged $\pi$, $K$ and $p$ at
    different collision energy in pp collisions.}
  \label{fig:pp-meanpt}
\end{figure}

Tsallis parameterization allows to find also the mean transverse
momentum $\langle \pT \rangle$ and to observe its evolution with
collision energy (Fig.\ref{fig:pp-meanpt}).
Comparison of mean $\pT$ of different hadron species measured at
different collision energies indicates that hadron production spectra
become harder at higher $\sqrt{s}$, and also mean $\pT$ grows with
hadron mass.

\begin{figure}[ht]
  \centering
  \includegraphics[width=0.48\hsize]{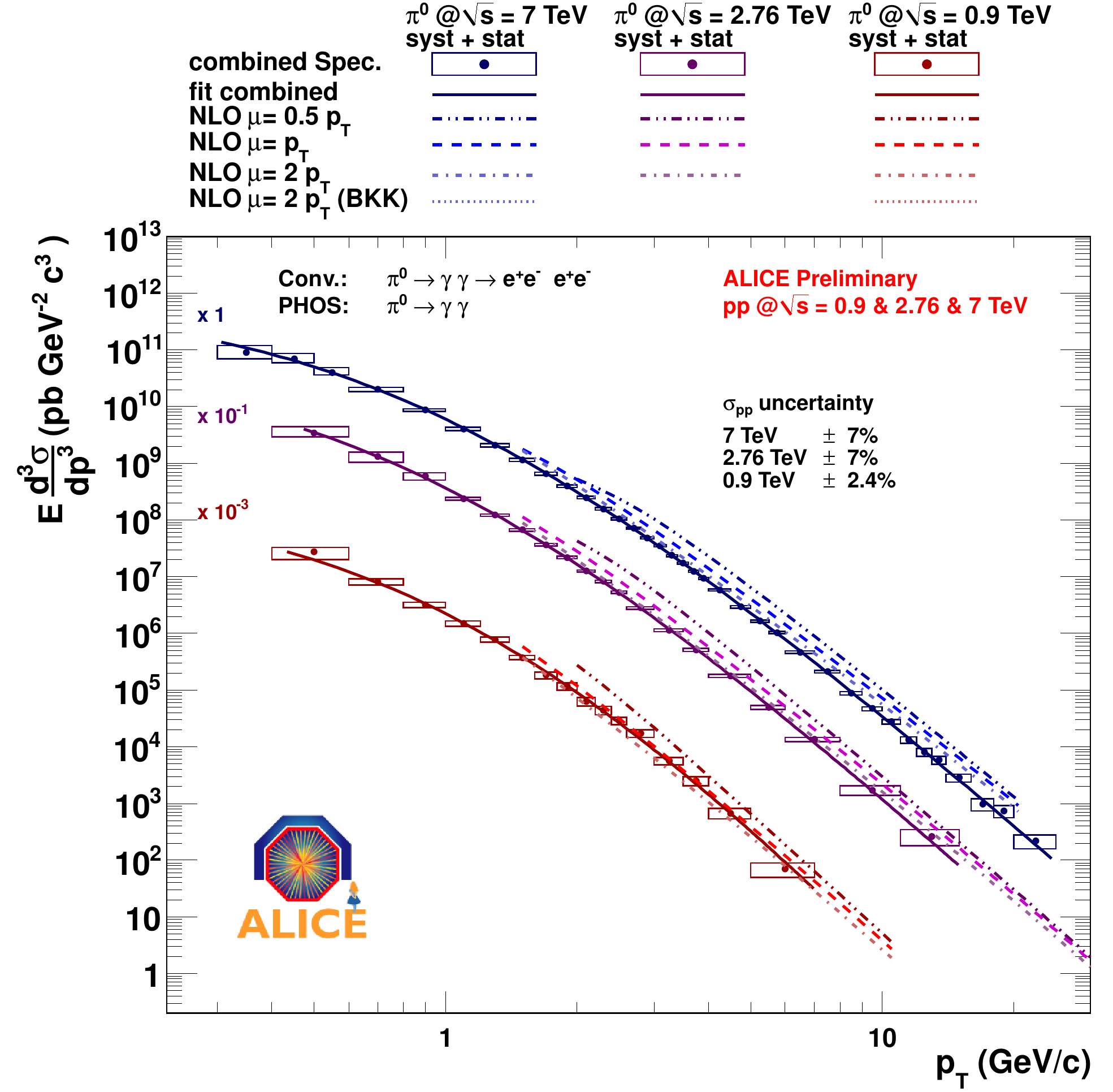}
  \includegraphics[width=0.48\hsize]{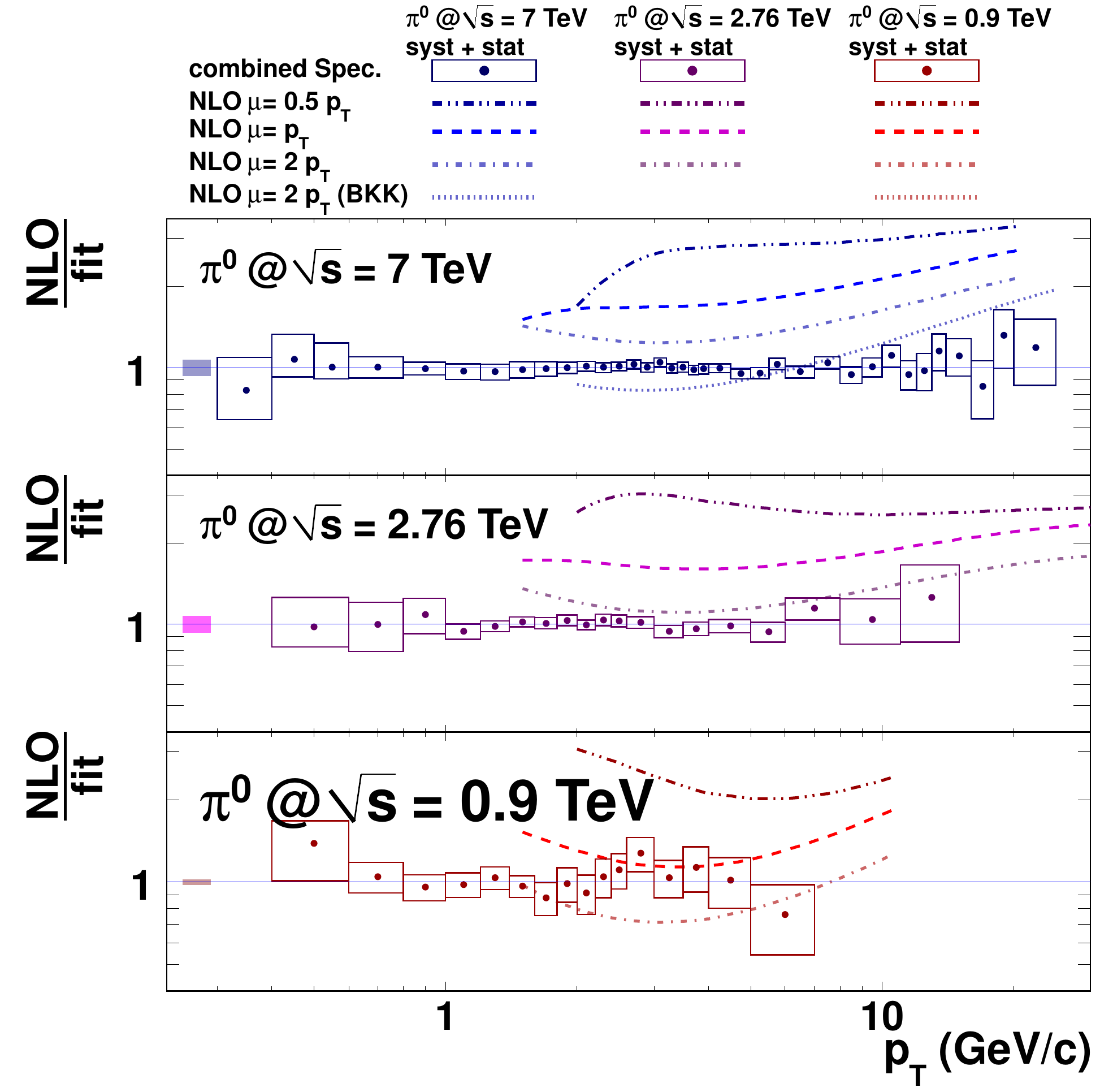}
  \caption{Production spectrum of $\pi^0$ in pp
    collisions at $\sqrt{s}=0.9$, $2.76$ and $7$~TeV (left) and
    ratio of NLO pQCD calculations to the measured spectra (right).}  
  \label{fig:pp-pi0}
\end{figure}

ALICE has also measured production spectra of neutral pions and $\eta$
mesons in pp collisions at $\sqrt{s}=0.9$, $2.76$ and 7~TeV, using the
Photon Spectrometer (PHOS) for real photon detection and central
tracking system for converted photon reconstruction
\cite{pp-pi0}. Neutral meson reconstruction, performed via invariant
mass spectra of photon pairs, allowed to measure differential cross
section of $\pi^0$ and $\eta$ in a wide \pT\ range. In particular, the
spectrum of $\pi^0$ production at the three collision energies are
shown of the left plot of Fig.\ref{fig:pp-pi0}.
Hadron production at high \pT\ can be well calculated in the
next-to-leading orders of perturbative QCD (NLO pQCD). These
calculations are based on parton distribution (PDF) and fragmentation
functions (FF) measured at lower energies. Application of those PDF's
and FF's to the new energy domain delivered by LHC, lead to
extrapolations of those functions to the kinematic region where the
functions have large uncertainties. The ratio of differential cross
sections of $\pi^0$ and $\eta$ mesons in pp collisions, calculated by
NLO pQCD, to the Tsallis fit of the ALICE measurements are shown by
curves on the right plot of Fig.\ref{fig:pp-pi0}. Data points on this
plot represent the ratio of the measured cross section to the Tsallis
fit to the measurement, which demonstrates the quality of the data
description by the Tsallis parameterization. This comparison of
theoretical calculations and experimental measurements demonstrates
that NLO pQCD at the QCD scale $\mu=\pT$ describes well hadron
production in pp collisions at $\sqrt{s}=0.9$~TeV, while significantly
overestimate it at $\sqrt{s}=7$~TeV. No common set of pQCD parameters
can be found to describe equally well the spectra of pion production
at all three collision energies.

\begin{figure}[ht]
  \includegraphics[width=0.48\hsize]{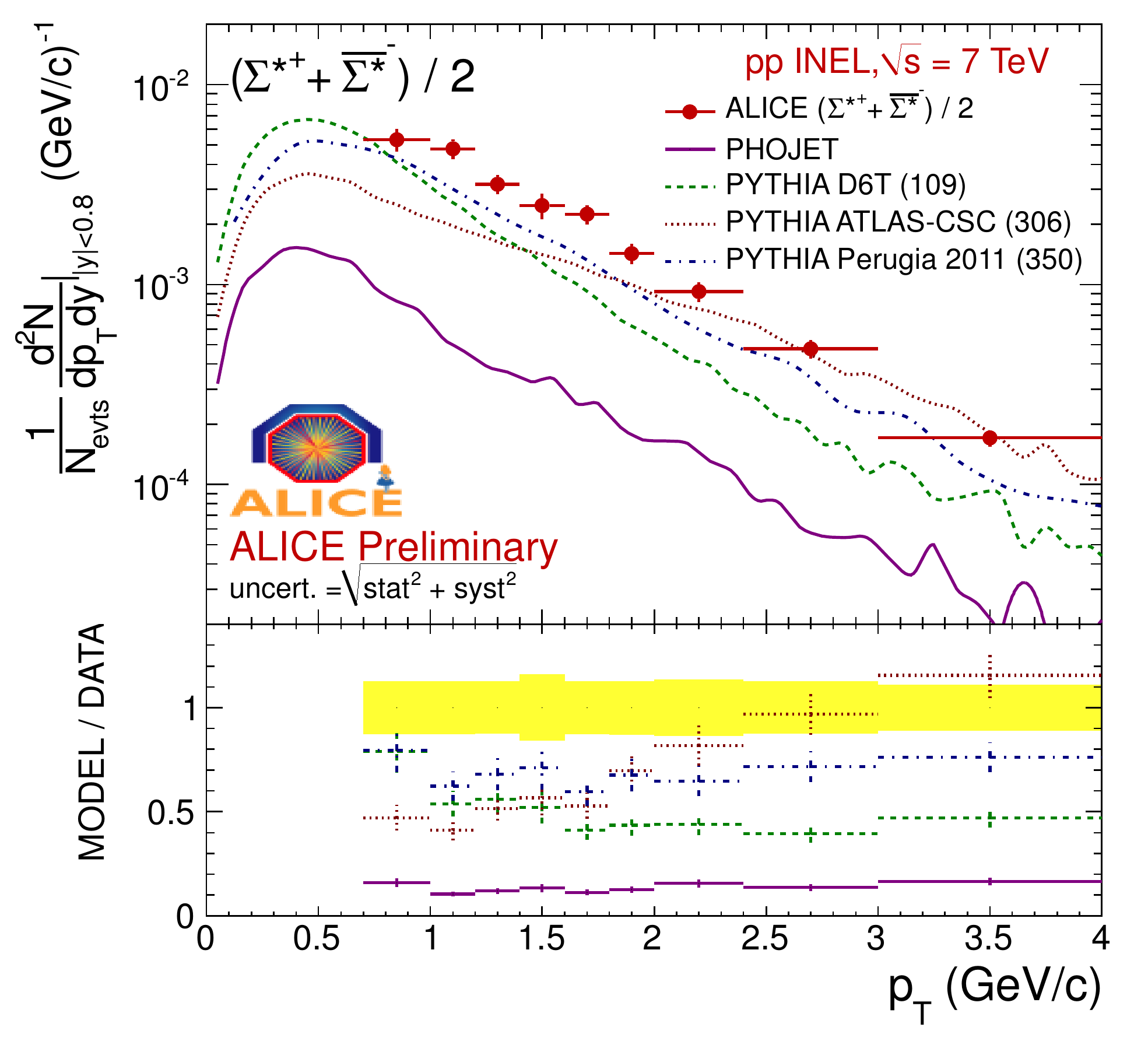}
  \hfil
  \includegraphics[width=0.48\hsize]{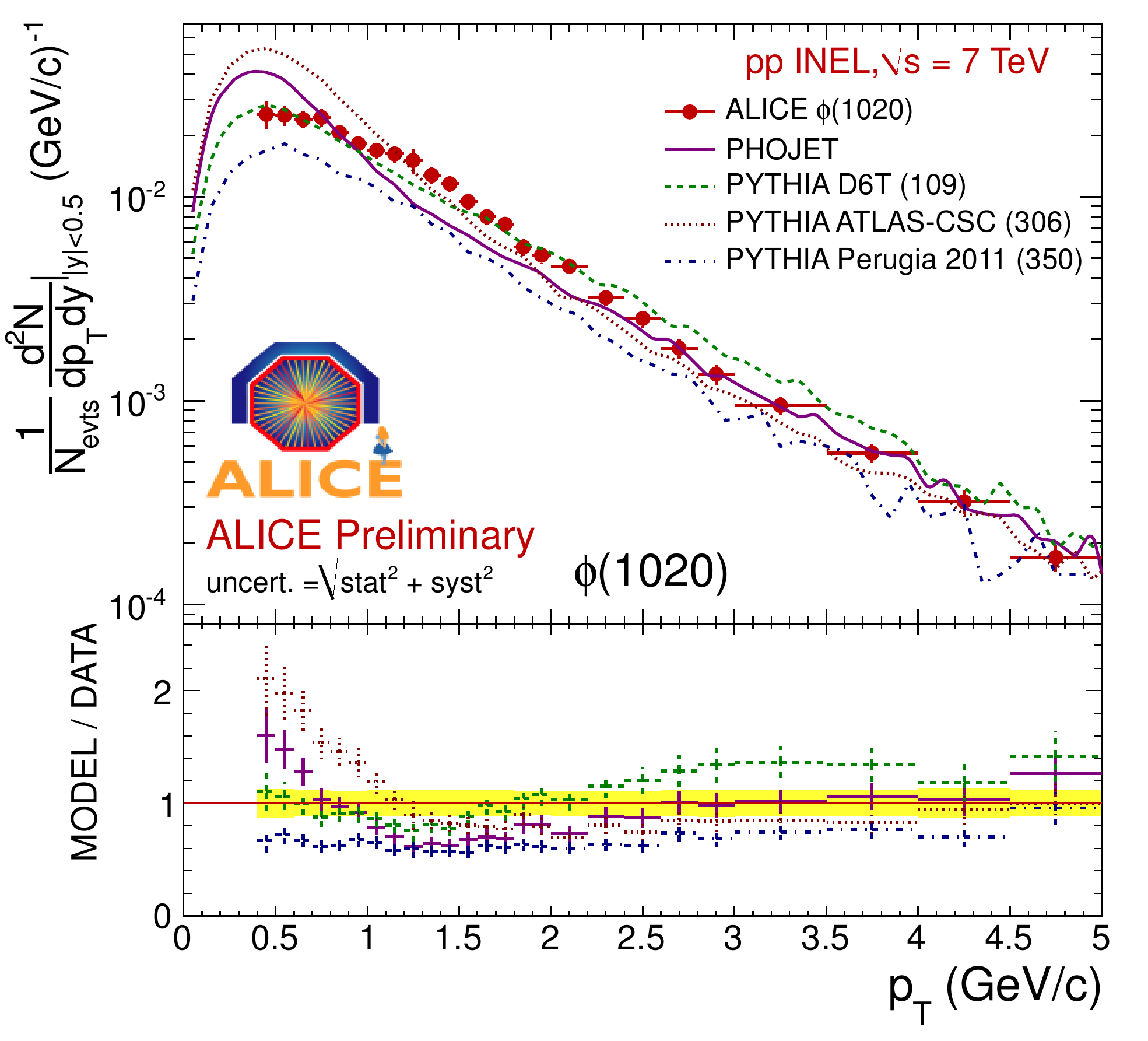}
  \caption{Production spectrum of $(\Sigma^{*+} + \bar{\Sigma^*}^-)/2$ and $\phi$ in pp
    collisions at $\sqrt{s}=7$~TeV with Monte Carlo
    predictions by different models.}  
  \label{fig:pp-resonances}
\end{figure}

Strangeness production is one of the most important observables for
studying the strongly interacting matter produced in heavy-ion
collisions. That is why measurements of complete set of strange
hadrons in pp collisions is necessary as a reference for comparison
with heavy ion collisions. Besides charged kaons mentioned earlier,
ALICE has measured production spectra of many other strange hadrons,
as well as those of mesons with hidden strangeness ($K^*$, $\Lambda$,
$\Sigma$, $\Omega$, $\phi$ and strange resonance baryons). Production
yields of $(\Sigma^*_{} + \bar{\Sigma^*}^-)/2$ and $\phi$ mesons in pp
collisions at $\sqrt{s}=7$~TeV are shown in
Fig.\ref{fig:pp-resonances} and are compared with several MC
predictions.

Identified hadron spectra measured at LHC energies, in conjunction
with spectra measured by previous experiments at lower collision
energies, allow to observe evolution of hadron production properties
with $\sqrt{s}$.  Predictions of various phenomenological models, as
well as NLO pQCD calculations were found to be unable to describe all
identified hadron spectra measured by ALICE in pp collisions

\section{Heavy ion collisions}

Analysis of the first heavy-ion data collected in 2010 brought many
results giving an insight into the properties of strongly interacting
matter at the new energy density regime. Observables characterizing
this matter are classified into several groups which will be reviewed
in this section.

\subsection{Global event properties}

\begin{figure}[ht]
  \includegraphics[width=0.51\hsize]{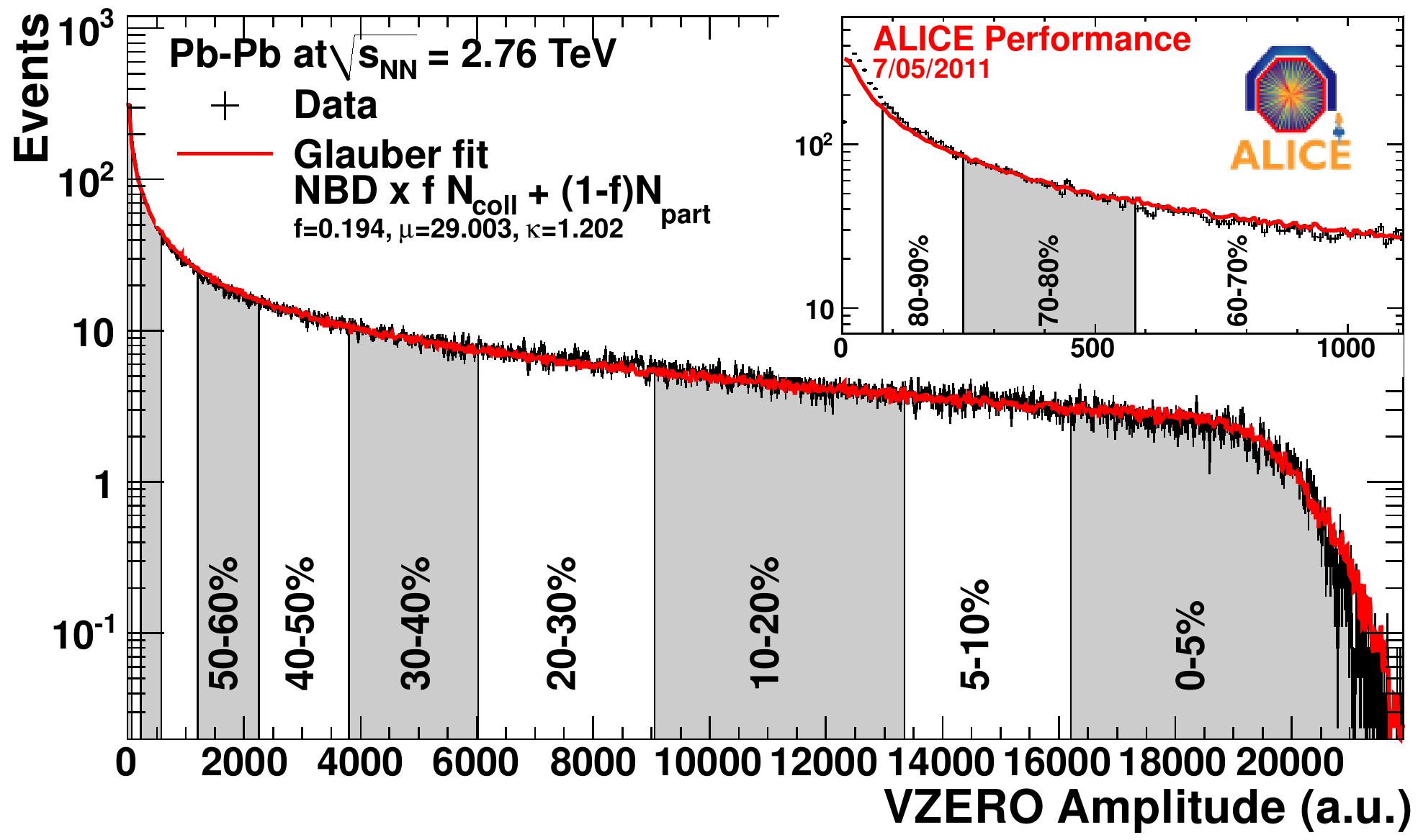}
  \hfill
  \includegraphics[width=0.46\hsize]{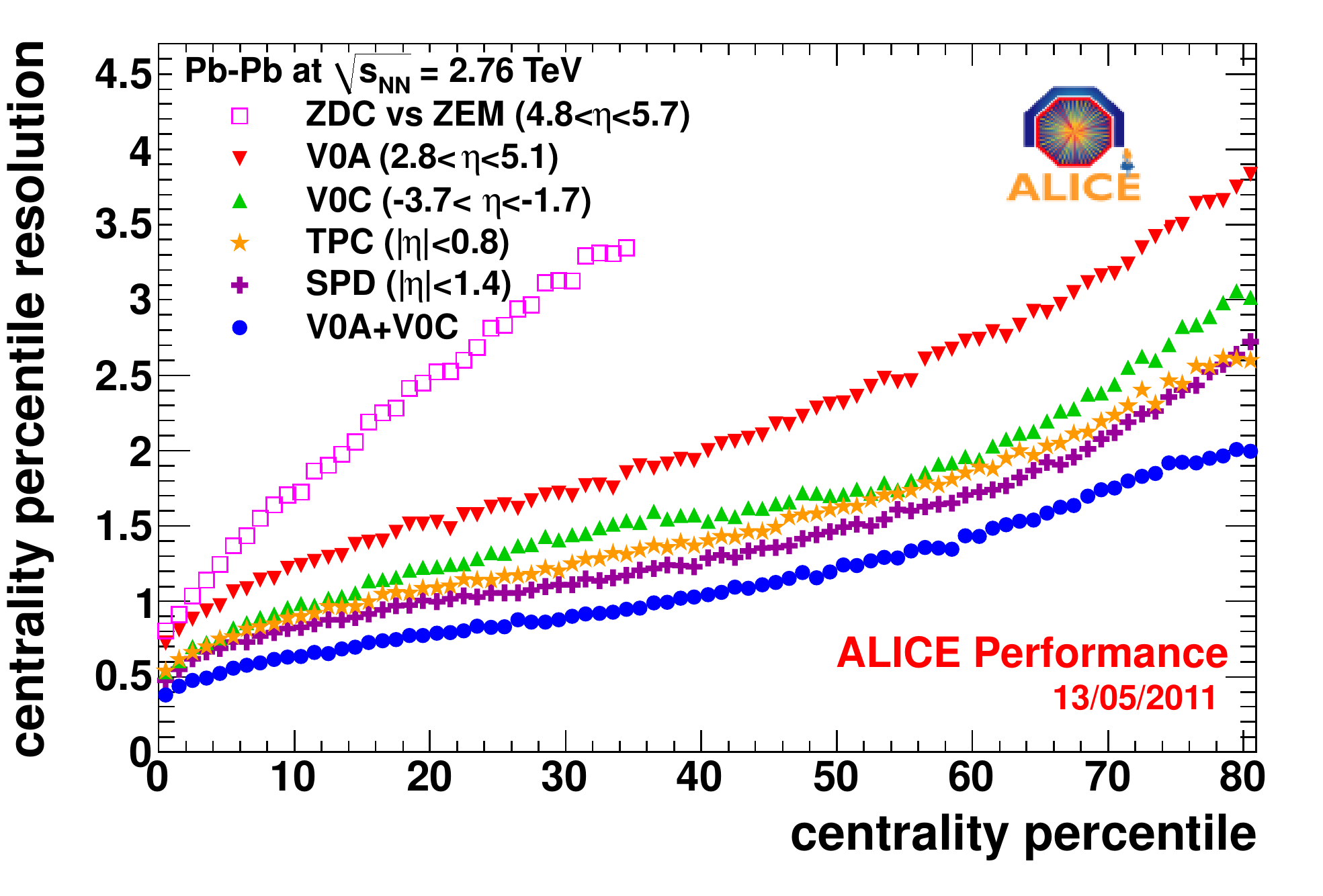}
  \caption{Centrality determination in ALICE. Glauber model fit to the
    VZERO amplitude with the inset of a zoom of the most peripheral
    region (left); Centrality resolution with different detectors
    (right).}
  \label{fig:PbPb-centrality}
\end{figure}

As heavy nuclei are extended objects, centrality determination is an
essential point for all heavy-ion measurements. Centrality of the
collision, directly related to the impact parameter and to the number
of nucleons $N_{\rm part}$ participating in the collision, allows to
study particle production versus the density of the colliding system.
In the ALICE experiment, collision centrality can be measured by
several detectors. The best accuracy of centrality measurement is
achieved with the scintillator hodoscope VZERO covering pseudorapidity
ranges $2.8 < \eta < 5.1$ and $-3.7 < \eta < -1.7$. Distribution of
the sum of amplitudes in VZERO in minimum bias Pb-Pb collisions is
shown in Fig.\ref{fig:PbPb-centrality} (left)
\cite{bib:PbPb-dNdy}. Centrality classes were defined by Glauber
model, and the fit of the Glauber model to the data is shown by a
solid line in this plot. Centrality resolution for all the estimators
can be found in Fig.\ref{fig:PbPb-centrality} (right)
\cite{bib:ToiaQM2011} which demonstrates that the best resolution is
achieved with the VZERO detector, and is equal to about 0.5\% in the
most central events, and varies up to 1.5\% in the most peripheral
collisions.

\begin{figure}[ht]
  \includegraphics[width=0.48\hsize]{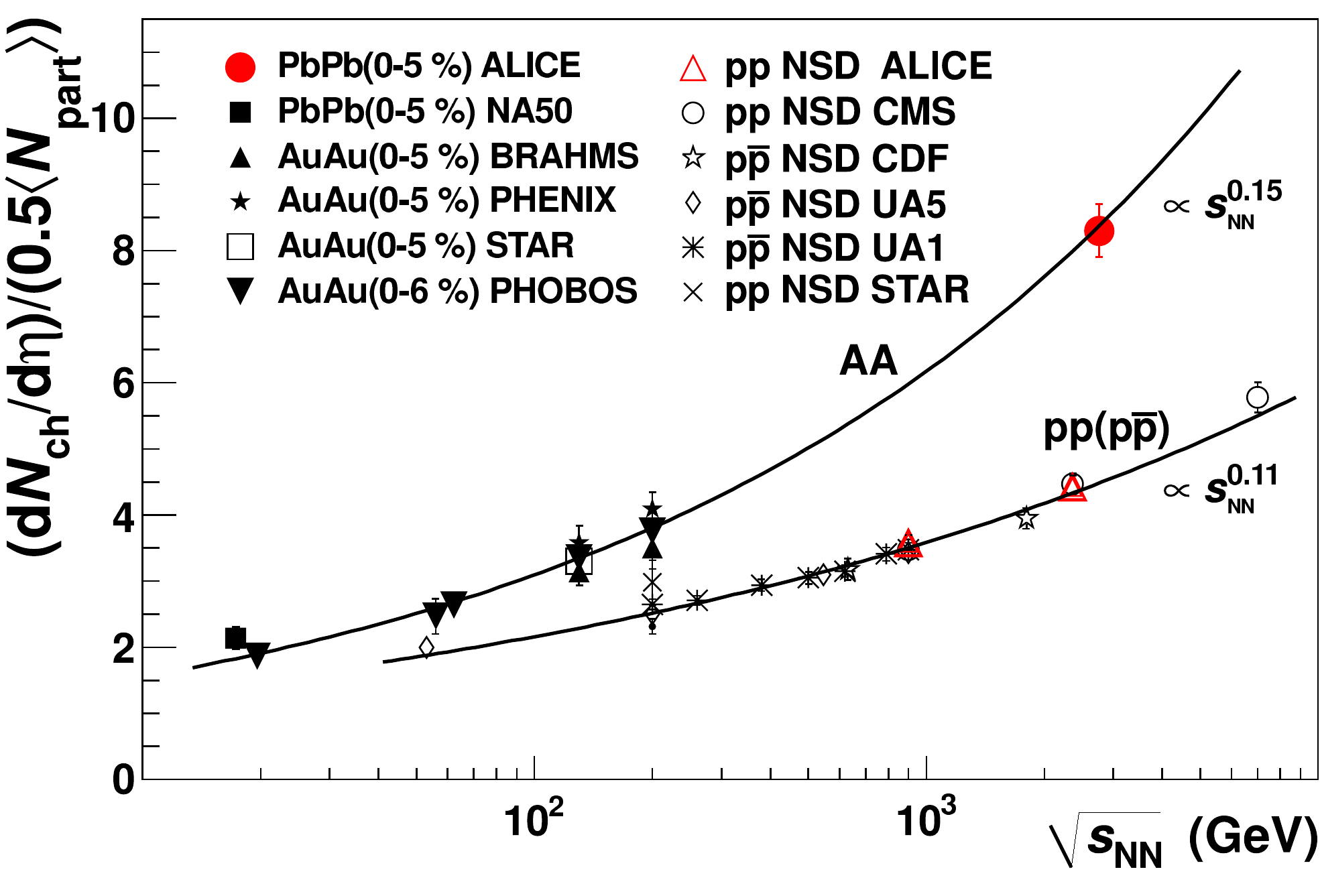}
  \hfill
  \includegraphics[width=0.48\hsize]{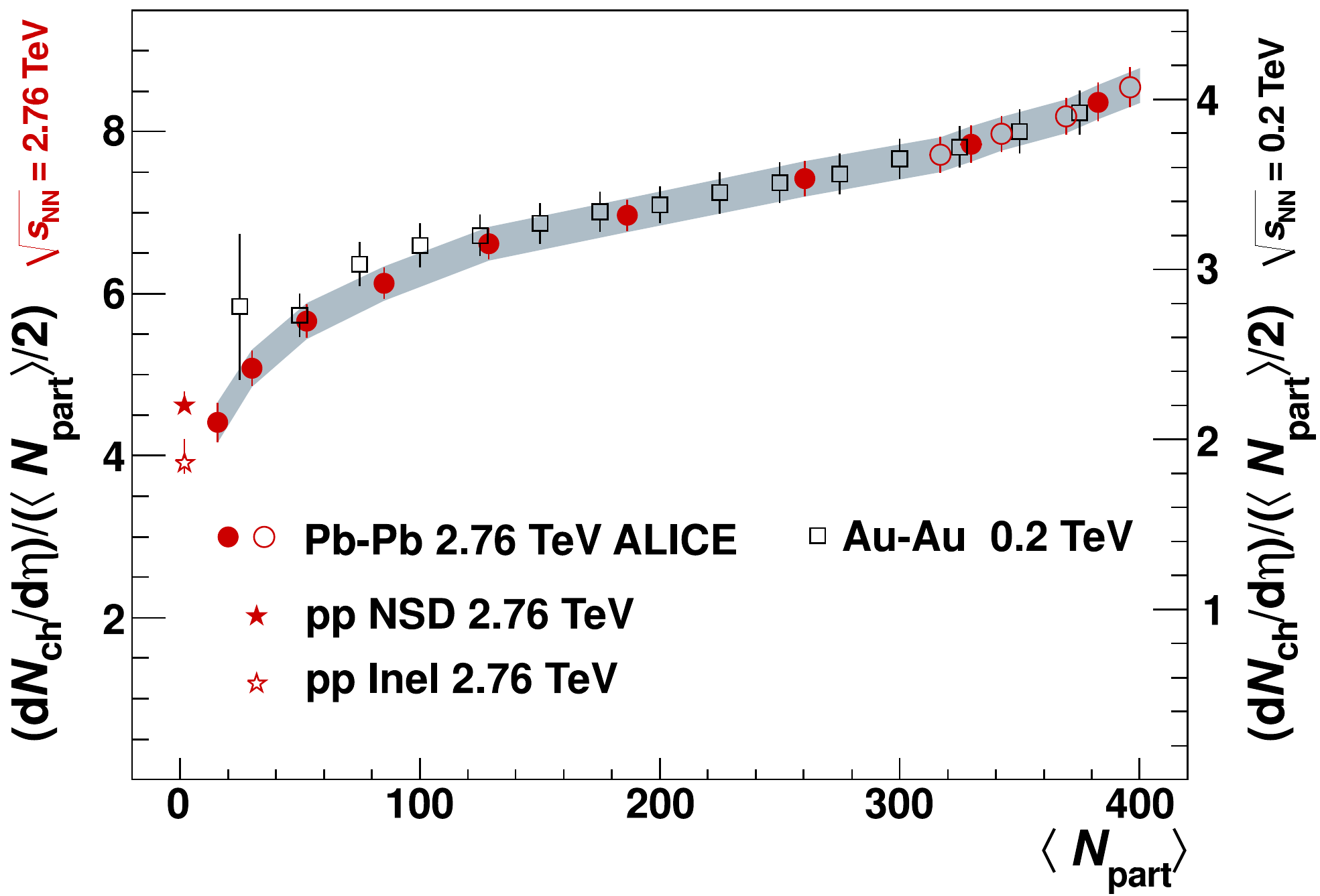}
  \caption{Charged track density $dN/d\eta$ in pp and AA collisions vs
    collision energy (left) and vs the number of participants (right).}
  \label{fig:PbPb-dNdeta}
\end{figure}

One of the key observable in heavy ion collision is the charged
particle multiplicity and its dependence on the collision
centrality. The main detector used for this measurements in the
Silicon Pixel Detector (SPD), two innermost layers of the barrel
tracking system covering the pseudorapidity range $|\eta|<1.4$. The
charged particle density, normalized to the average number of
participants in a given centrality class, $dN_{\rm
  ch}/d\eta/\left(\langle N_{\rm part} \rangle \right)$ was measured
by ALICE in PbPb collisions at $\sqsNN=2.76$~TeV and compared with
similar measurements at lower energies at RHIC and SPS
(Fig.\ref{fig:PbPb-dNdeta}, left plot) \cite{bib:ToiaQM2011}. 
In the most central events (centrality $0-5\%$) at LHC energy the
charged particle density was found to be $dN_{\rm ch}/d\eta=1601\pm
60$ \cite{bib:PbPb-dNdy} which is, being normalized to the number of
participants, is 2.1 times larger than the charged particle density
measured at RHIC at $\sqsNN=200$~GeV and 1.9 times larger than that in
pp collisions at $\sqrt{s}=2.36$~TeV. The dependence of $dN_{\rm
  ch}/d\eta$ on the number of participants $N_{\rm part}$, shown in the
right plot of Fig.\ref{fig:PbPb-dNdeta}, is very similar at LHC
($\sqsNN=2.76$~TeV) and RHIC ($\sqsNN=0.2$~TeV) energies, provided
the RHIC points are scaled by a factor 2.1 to match the LHC points.

\begin{figure}[ht]
  \includegraphics[width=0.48\hsize]{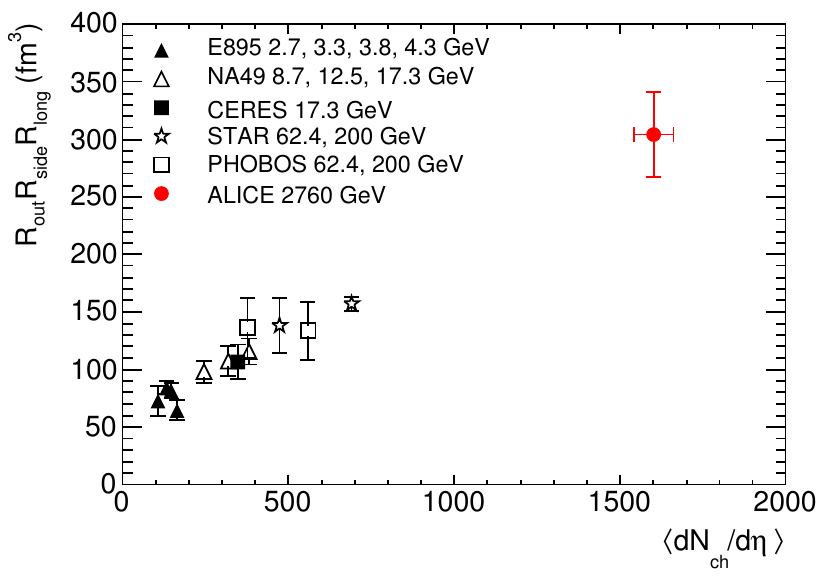}
  \hfill
  \includegraphics[width=0.48\hsize]{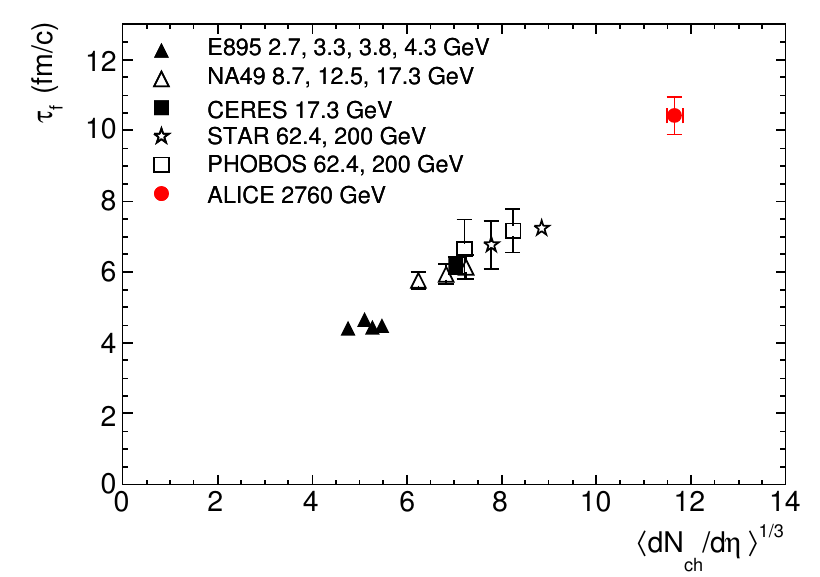}
  \caption{System size (left) and lifetime (right).}
  \label{fig:PbPb-HBT}
\end{figure}

Longitudinal and transverse expansion of the highly compressed
strongly-interacting system created in heavy-ion collisions can be
studied experimentally via intensity interferometry, the Bose-Einstein
enhancement of identical bosons emitted close by in phase space, known
as Hanbury Brown-Twiss analysis (HBT). ALICE has measured the HBT
radii and evaluate space-time propertied on the system generated in
Pb-Pb collisions at $\sqsNN=2.76$~TeV \cite{bib:HBT}. The two-particle
correlation function of the difference $\vec{q}$ of two 3-momenta
$\vec{p_1}$ and $\vec{p_2}$ was measured for like-sign charged pions
which allowed to get the Gaussian HBT radii, $R_{\rm out}$, $R_{\rm
  side}$ and $R_{\rm long}$. The product of these 3 radii and
decoupling time extracted from $R_{\rm long}$, measured by
ALICE at LHC energy, together with this value measured at the AGS, SPS
and RHIC, is shown in Fig.\ref{fig:PbPb-HBT} (left) as a function of
charged track density $dN_{\rm ch}/d\eta$.
This measurements indicate that the homogeneity volume in central PbPb
collisions at $\sqsNN=2.76$~TeV exceeds that measured at RHIC by a
factor of 2. The increase is present in both longitudinal and
transverse radii. The decoupling time for mid-rapidity pions exceeds 10
fm/c which is 40\% larger than at RHIC (Fig.\ref{fig:PbPb-HBT}, right).

\subsection{Collective expansion}

In non-central collision of nuclei, the overlap region, and hence the
initial matter distribution is anisotropic. During evolution of the
matter, the spatial asymmetry of initial state is converted to an
anisotropic momentum distribution. The azimuthal distribution of the
particle yield can be expressed in terms of the angle between the
particle direction $\varphi$ and the reaction place $\Psi_{\rm RP}$:
\begin{eqnarray}
\frac{dN}{d(\varphi-\Psi_{\rm RP})} & \propto
& 1+2\sum_{n=1}^{}v_n \cos\left[n(\varphi-\Psi_{\rm RP})\right], \\
& & v_2 = \langle \cos\left[n(\varphi-\Psi_{\rm RP})\right] \rangle.
\end{eqnarray}
The second coefficient of this Fourier series, $v_2$, is referred to
as elliptic flow. Theoretical models, based on relativistic
hydrodynamics \cite{bib:hydro-v2_Kestin,bib:hydro-v2_Niemi},
successfully described the elliptic flow observed at RHIC
\cite{bib:RHIC_v2} and predict its increase at LHC energies from 10\%
to 30\%.

\begin{figure}[ht]
  \includegraphics[width=0.52\hsize]{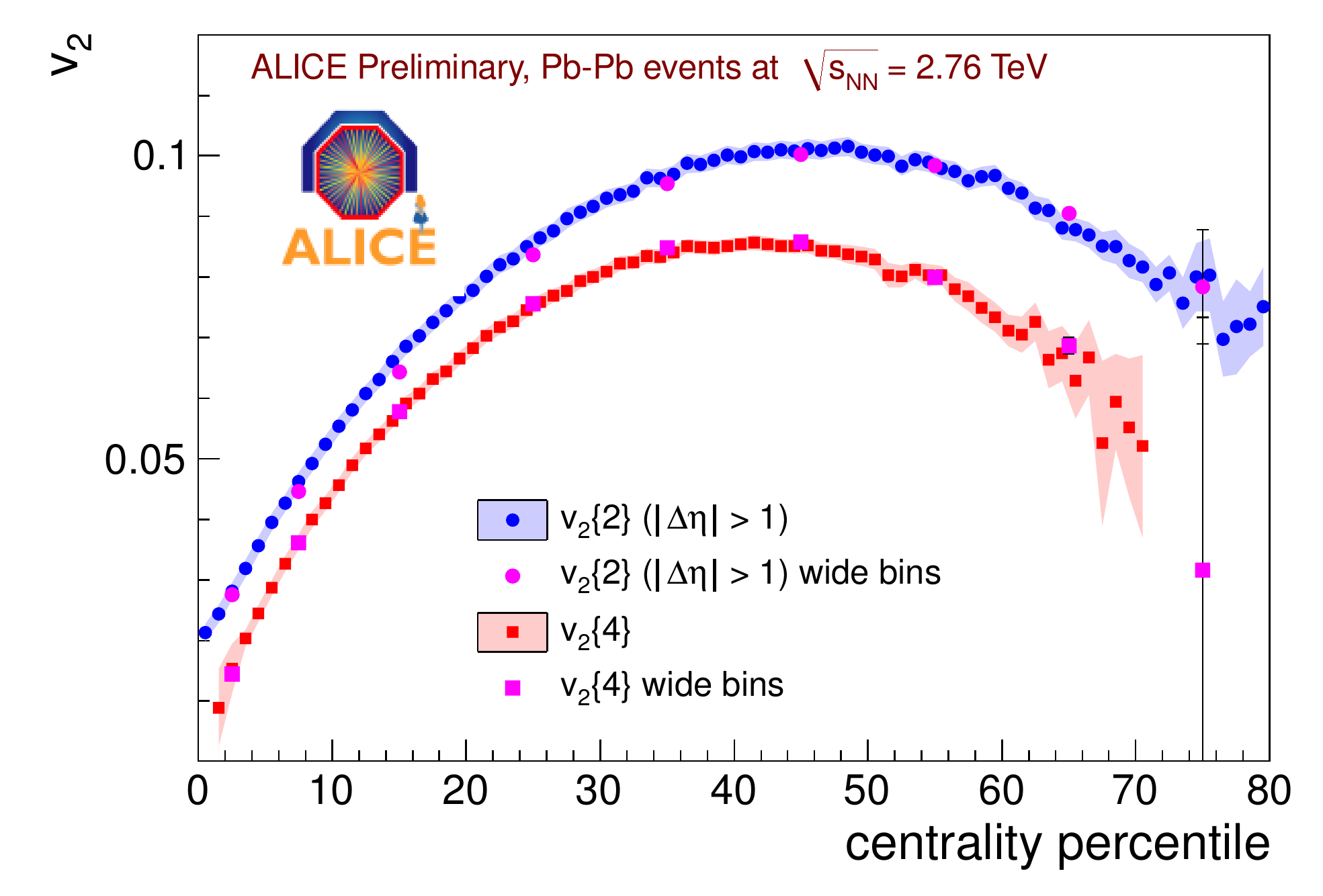}
  \hfill
  \includegraphics[width=0.44\hsize]{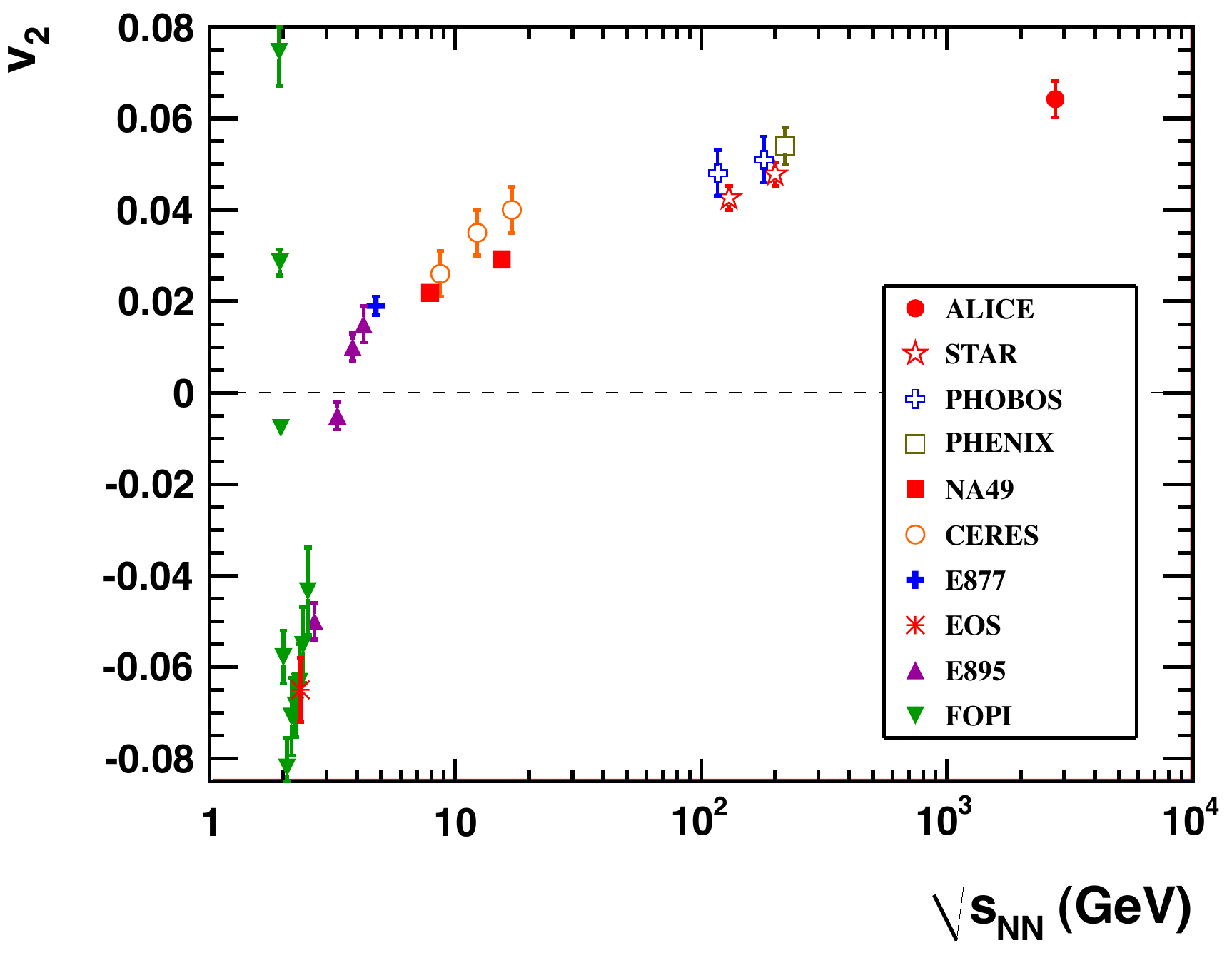}
  \caption{Azimuthal flow $v_2$ of charged particles in Pb-Pb
    collisions at $\sqsNN=2.76$~TeV vs centrality (left) and
    $v_2$ vs collision energy (right).}
  \label{fig:PbPb-v2}
\end{figure}

The first measurements of elliptic flow of charged particles in Pb-Pb
collisions at $\sqsNN=2.76$~TeV were reported by ALICE in
\cite{bib:ALICE-v2}. Charged tracks were detected and reconstructed in
the central barrel tracking system, consisting of ITS and
TPC. Elliptic flow integrated over $\pT$ range $0.2 < \pT <
5$~GeV/$c$, for the 2- and 4-particle cumulant methods, is shown in
Fig.\ref{fig:PbPb-v2} (left) as a function of centrality.
It shows that the integrated elliptic flow increases from central to
peripheral collision and reaches the maximum value $v_2 \approx 0.1$
in semi-central collisions in the $40-60\%$ centrality
class. Comparison of the integrated elliptic flow of charged
particles, measured at different center-mass collision energies, shows
a smooth increase of $v_2$ with $\sqsNN$, and confirms model
expectations that the value of $v_2$ in Pb-Pb collisions at
$\sqsNN=2.76$~TeV increases by about 30\% with respect to $v_2$ in
Au-Au collisions at $\sqsNN=0.2$~TeV.

\begin{figure}[ht]
  \includegraphics[width=0.48\hsize]{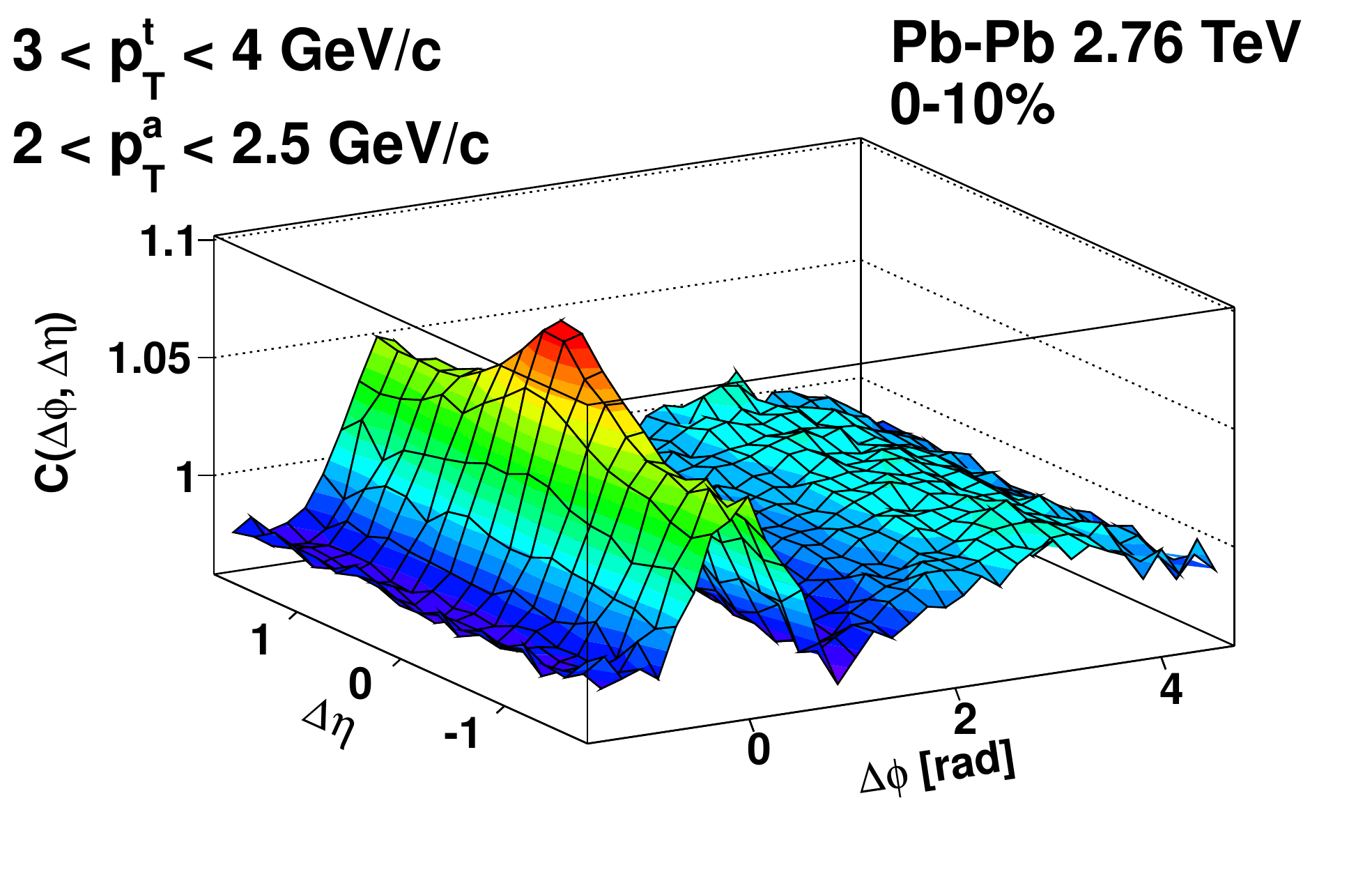}
  \hfill
  \includegraphics[width=0.48\hsize]{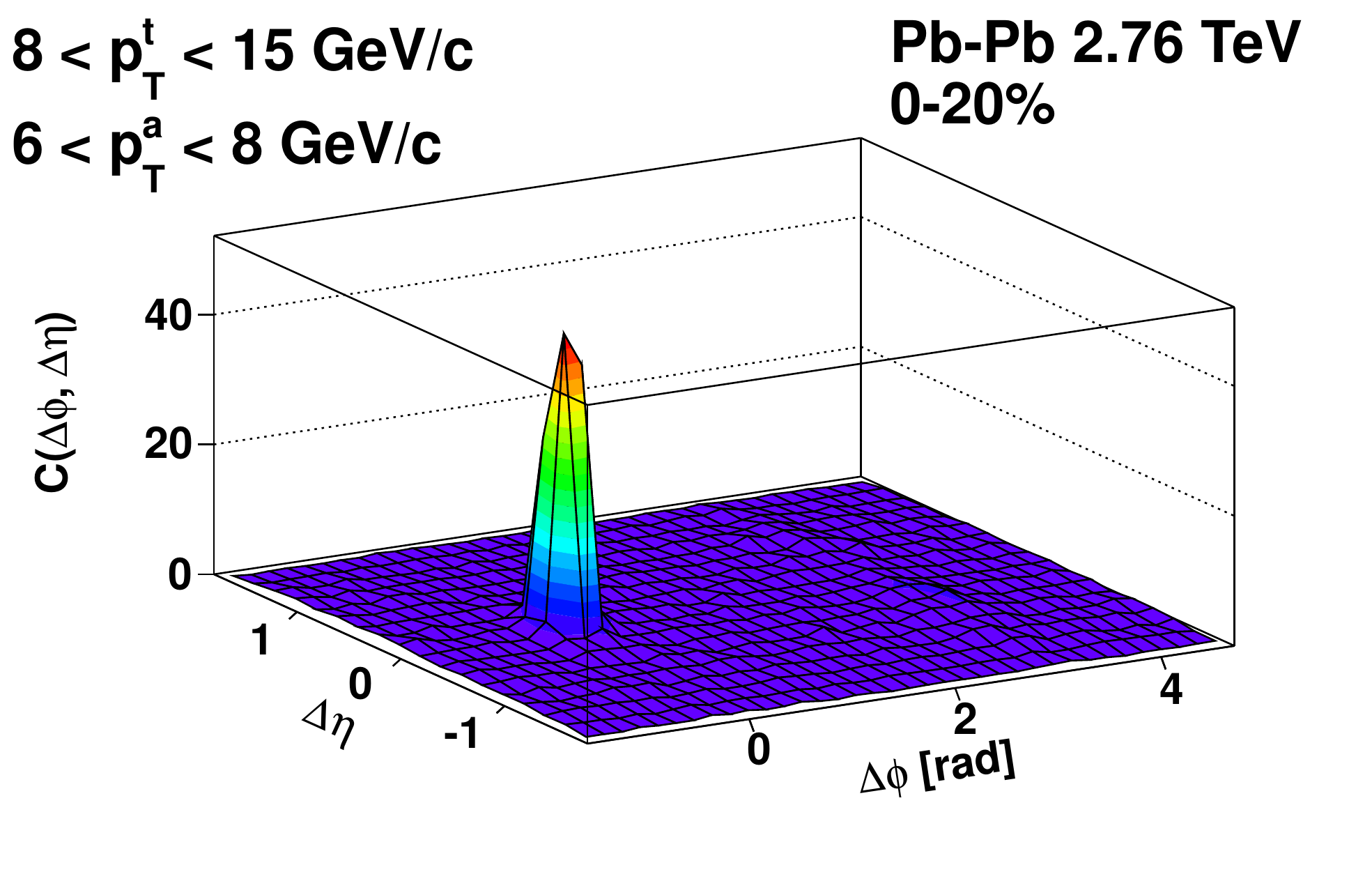}
  \caption{Di-hadron correlations $C(\Delta\varphi,\Delta\eta)$ in
    central Pb-Pb collisions in the ``bult-dominated'' regime (left)
    and in the ``jet-dominated'' regime (right).}
  \label{fig:PbPb-ridge}
\end{figure}

Particle momentum anisotropy is also studied via two-particle
correlations which measure the distributions of azimuthal angles
$\Delta\varphi$ and pseudorapidities $\Delta\eta$ between a
``trigger'' particle at transverse momentum $\pT^t$ and an
``associated'' particle at $\pT^a$. The correlation function
$C(\Delta\varphi,\Delta\eta)$ looks differently in different kinematic
regions. At $\pT^t < 3-4$~GeV/$c$, the shape of the correlation function
reveals the ``bulk-dominated'' regime, where hydrodynamic
modeling has been demonstrated to give a good description of the data
from heavy-ion collisions (see Fig.\ref{fig:PbPb-ridge}, left).
At high transverse momenta of both particles, jets become dominating,
and the shape of the correlation function in central Pb-Pb collisions
has just a clear near-side peak centered at $\Delta\varphi =
\Delta\eta = 0$ and no evident out-side peak, as shown in
Fig.\ref{fig:PbPb-ridge}, right. Harmonic decomposition of
two-particle correlations \cite{bib:ALICE-harmonic} performed by
ALICE, has shown that in the ``bulk-dominated'' regime a distinct
near-side ridge and a doubly-peaked away-side structure are observed
in the most central events, which reflects a collective response to
anisotropic initial conditions.

The results of global event properties and collective expantion
studied by ALICE, indicate that the fireball formed in nuclear
collisions at the LHC is hotter, lives longer, and expands to a larger
size at freeze-out as compared to lower energies.

\subsection{Strangeness production}

\begin{figure}[ht]
  \includegraphics[width=0.48\hsize]{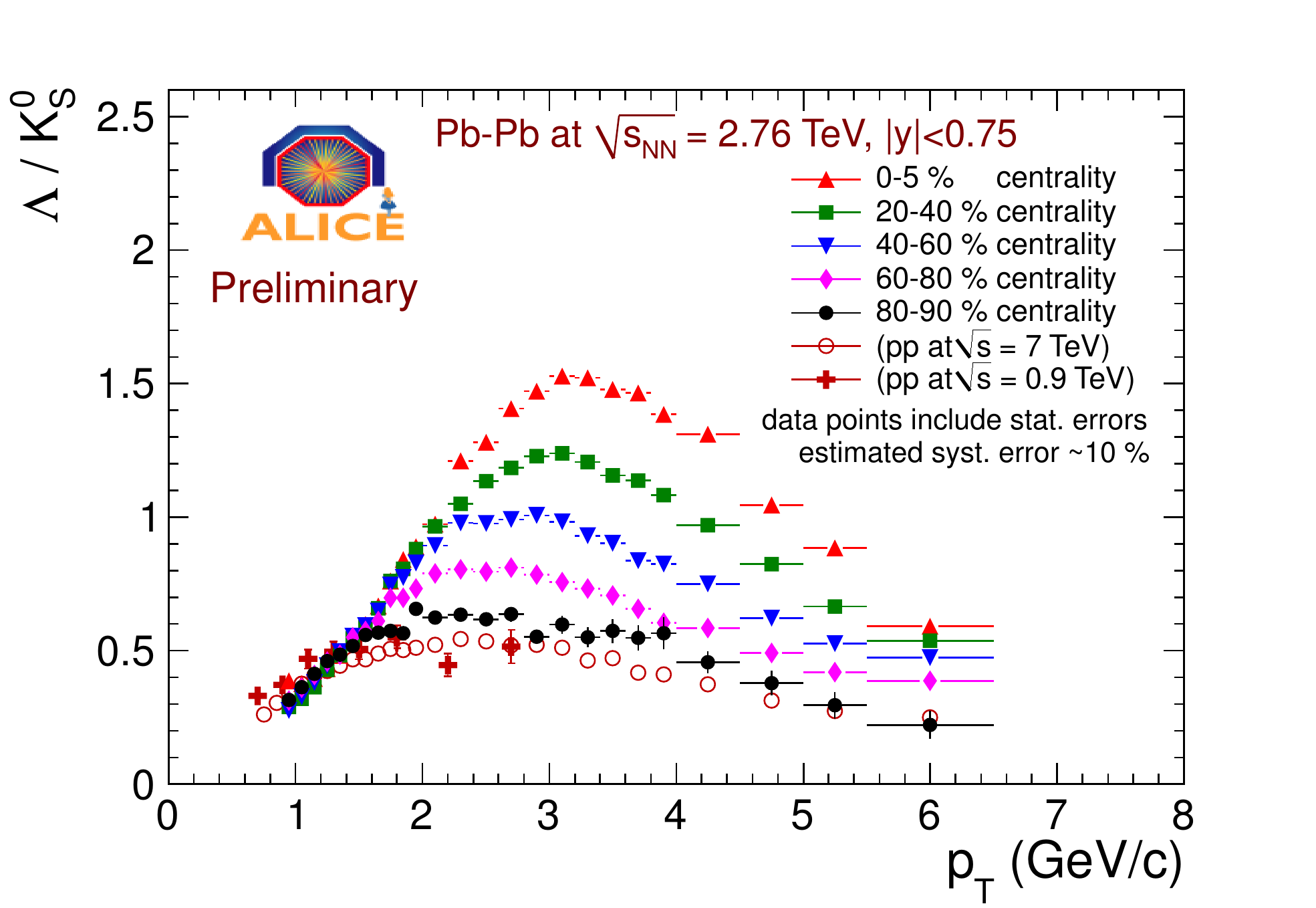}
  \hfill
  \includegraphics[width=0.48\hsize]{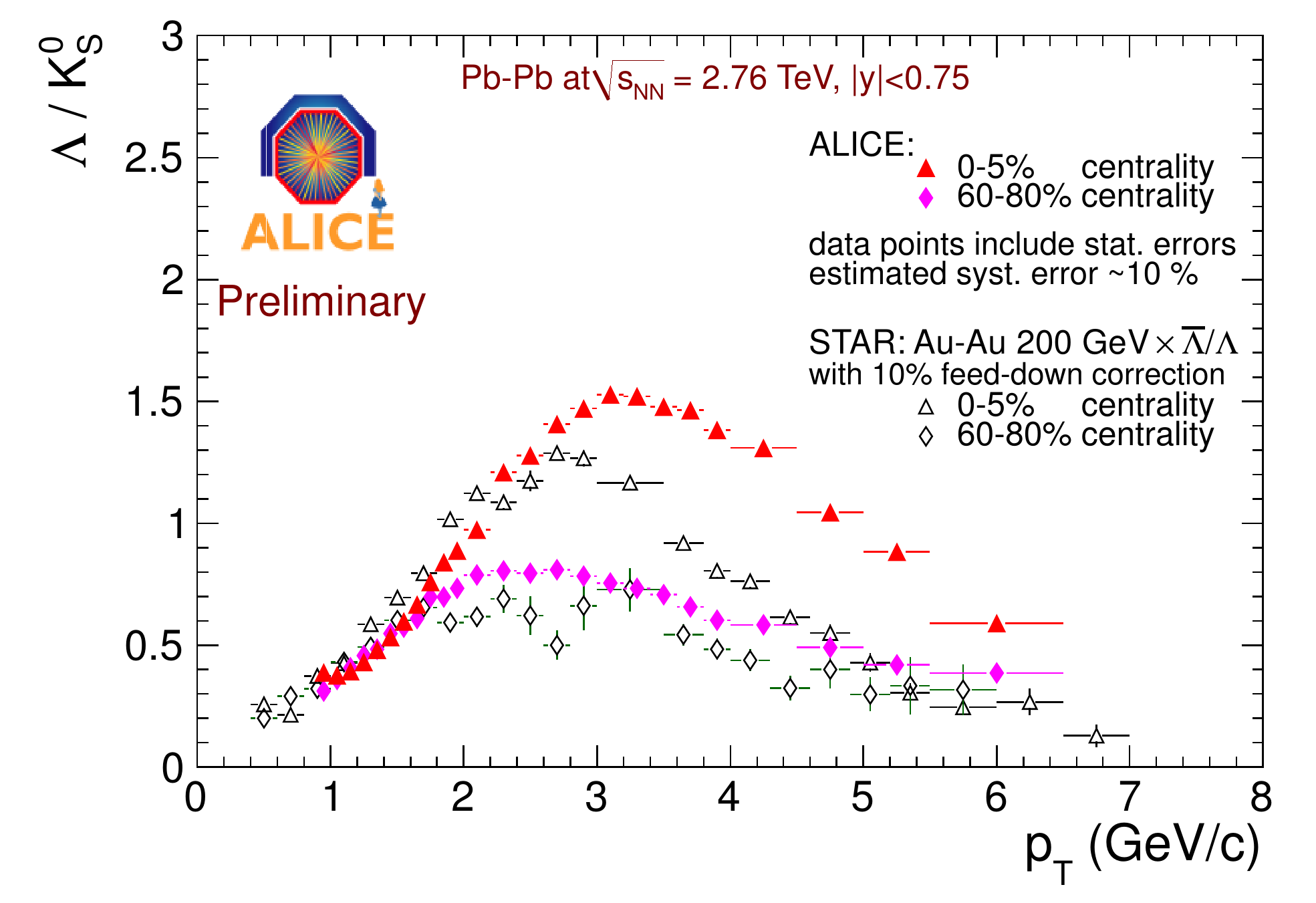}
  \caption{Ratio $\Lambda/K^0_S$ in Pb-Pb collisions at
    $\sqsNN=2.76$~TeV in different centralities (left) and
    comparison of this ratio at LHC and RHIC in centralities $0-5\%$
    and $60-80\%$ (right).}
  \label{fig:PbPb-Lambda_K0s}
\end{figure}

Strange particle production has been considered as a probe of strongly
interacting matter by heavy-ion experiments at AGS, SPS and RHIC. We
have already demonstrated that ALICE, due to its powerful particle
identification technique, has measured strange particle spectra in pp
collisions. Similar analysis was performed on the Pb-Pb data collected
in 2010. Comparison of strange meson and baryon production is
illustrated by the $\Lambda/K^0_S$ ratio measured by ALICE in
different centrality classes (Fig.\ref{fig:PbPb-Lambda_K0s},
left). This ratio in peripheral Pb-Pb collision is similar to that one
measured in pp collisions, but it grows with centrality,
increasing the value of 1.5 in the most central collisions. The
qualitative behaviour of this ratio on $\pT$ at the LHC collision
energy is similar to the ratio measured at RHIC by the STAR experiment
(Fig.\ref{fig:PbPb-Lambda_K0s}, right). An enhancement of strange and
multi-strange baryons ($\Omega^-$, $\bar{\Omega}^+$,
$\Sigma^-$,$\bar{\Sigma}^+$ ) was obsevred in heavy-ion collisions by
experiments at lower energies, and was confirmed by ALICE at LHC
energy \cite{ALICE-Hippolyte}. It was also shown that multi-strange
baryon enhancement scales with the number of participants $N_{\rm
  part}$ and decreases with the collision energy.

\subsection{Parton energy loss in medium}

Experiments at RHIC reported that hadron production at high
transverse momentum in central Au-Au collisions at a center-of-mass
energy per nucleon pair $\sqsNN=200$~GeV is suppressed by a factor
$4-5$ compared to expectations from an independent superposition of
nucleon-nucleon collisions. This suppression is attributed to energy
loss of hard partons as they propagate through the hot and dence QCD
medium. Therefore, a spectrum suppression of hadron production can be
used as a measure of the properties of the strongly interacting matter.

The strength of suppression of a hadron $h$ is expressed by
the nuclear modification factor $R_{AA}$, defined as a ratio of the
particle spectrum in heavy-ion collision to that in pp, scaled by the
number of binary nucleon-nucleron collisions $N_{\rm coll}$:
\begin{equation}
R_{AA}(\pT) = \frac{(1/N_{AA})d^2N_h^{AA}/d\pT d\eta}
                  {N_{\rm coll}(1/N_{pp})d^2N_h^{pp}/d\pT d\eta}.
\end{equation}
At the larger LHC energy, the density of the medium is expected
to be higher than at RHIC, leading to a larger energy loss of
high-$\pT$ partons. However, the hadron production spectra are less
steeply falling with $\pT$ at LHC than at RHIC which would reduce the
value of $R_{AA}$ for a given value of the parton energy loss.

\begin{figure}[ht]
  \centering
  \includegraphics[width=0.50\hsize]{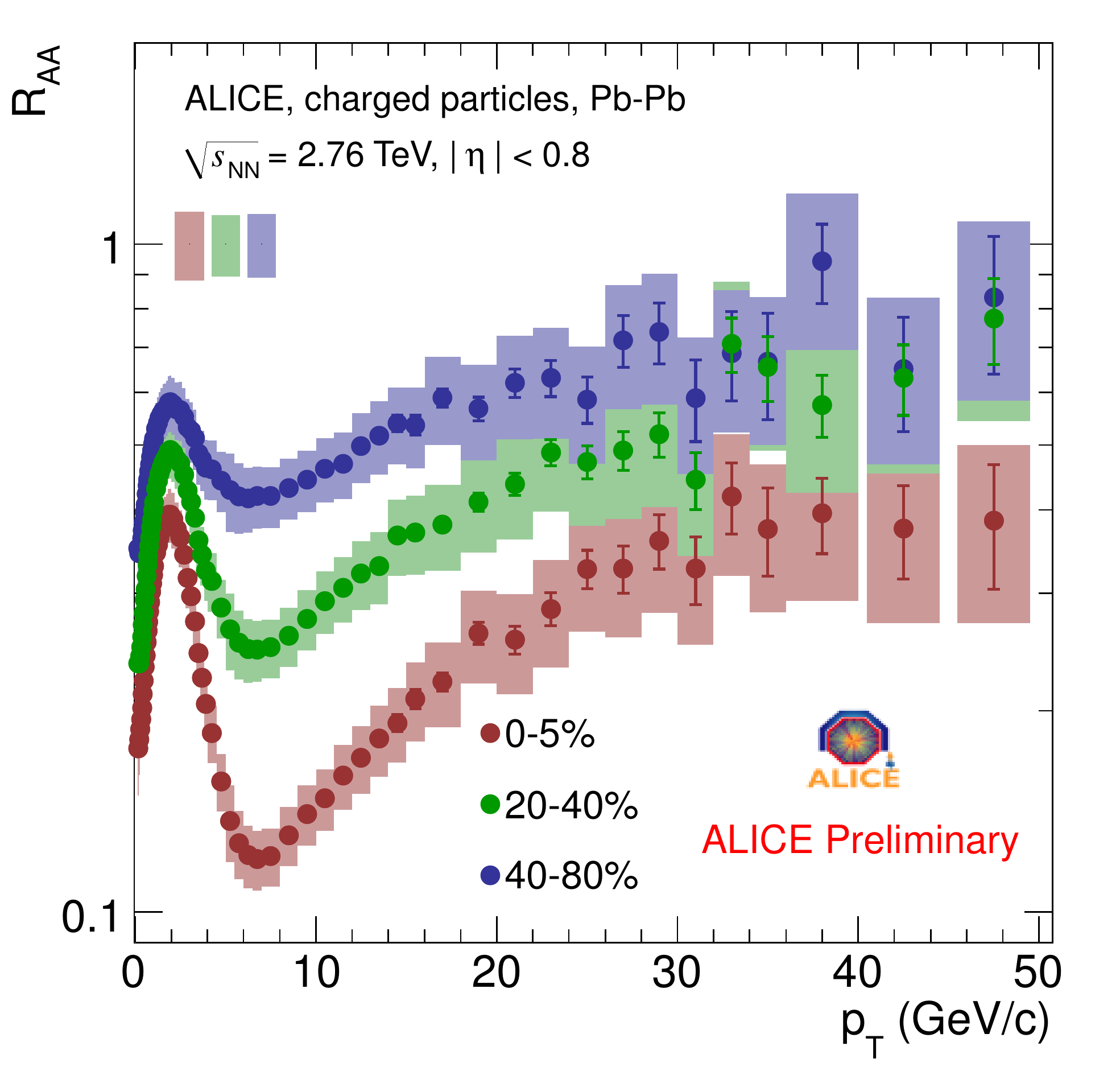}
  \caption{Nuclear modification factor $R_{AA}$ of charged particles.}
  \label{fig:PbPb-RAA_charged}
\end{figure}

ALICE has measured the nuclear modification factor $R_{AA}$ for many
particles. All charged particles, detected in the ALICE central
tracking system (ITS and TPC), show a spectrum suppression
\cite{Otwinowski:2011gq} which is qualitatively similar to that
observed at RHIC (Fig.\ref{fig:PbPb-RAA_charged}). However,
quantitative comparison with RHIC demonstrates that the suppression at
LHC energy is stronger which can be interpreted by a denser medium.
Benefiting from particle identification which has been already mention
earlier in this paper, ALICE has measured suppression of various
identified hadrons, which provides experimental data for studying the
flavor and mass dependence of the spectra suppression.

\begin{figure}[ht]
  \includegraphics[width=0.60\hsize]{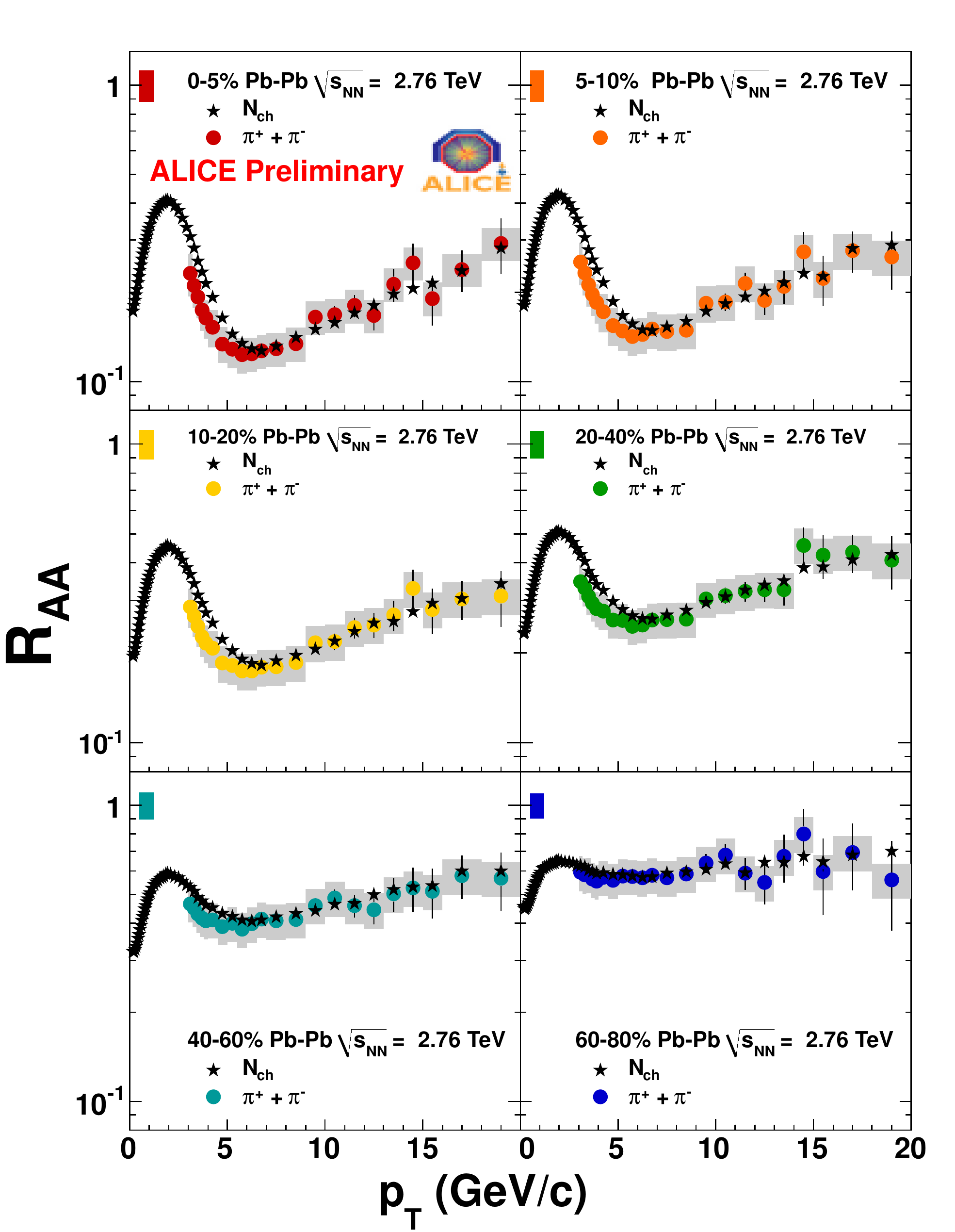}
  \caption{Nuclear modification factor $R_{AA}$ of charged pions.}
  \label{fig:PbPb-RAA_pions}
\end{figure}

\begin{figure}[ht]
  \includegraphics[width=0.48\hsize]{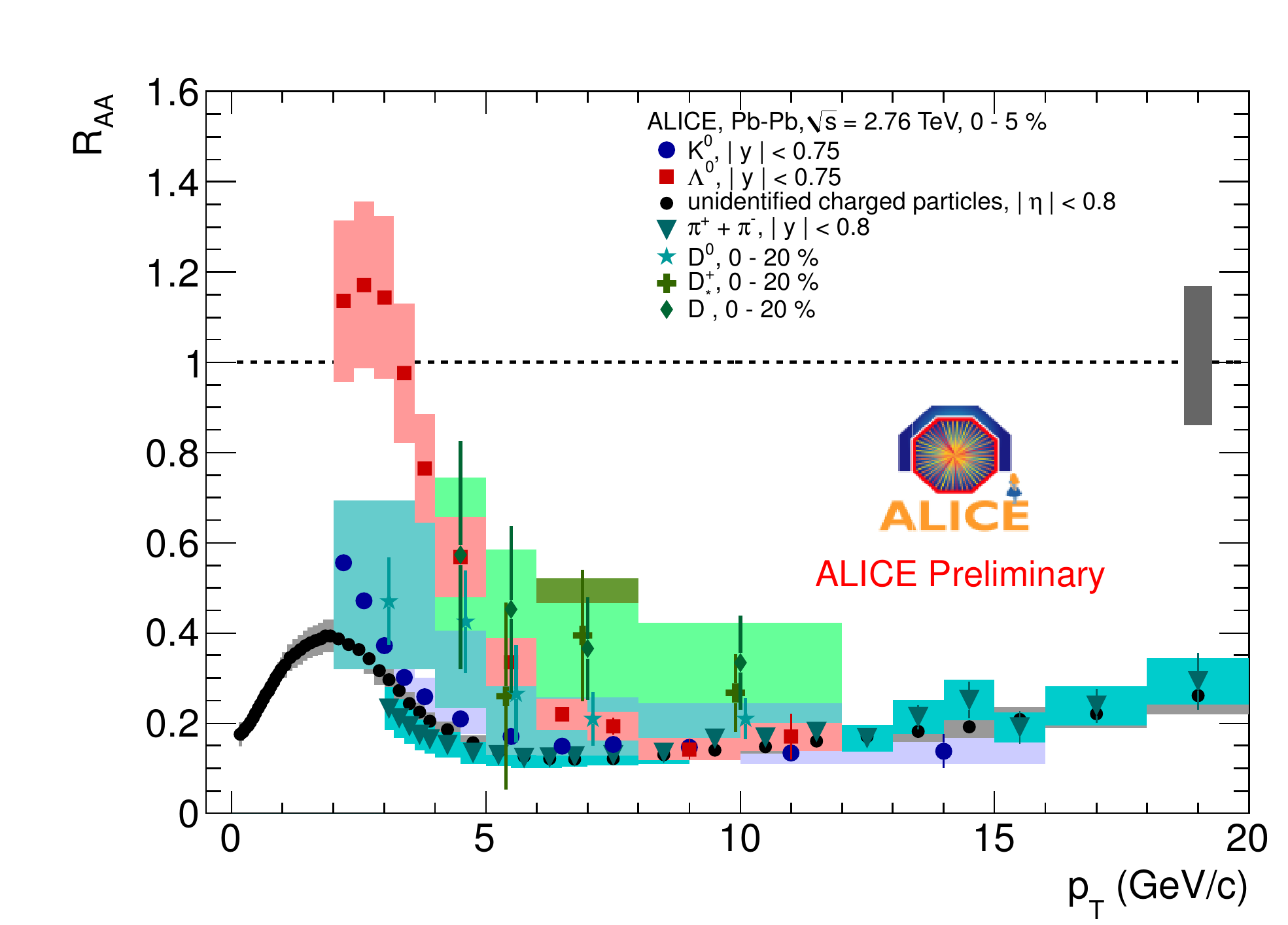}
  \hfill
  \includegraphics[width=0.48\hsize]{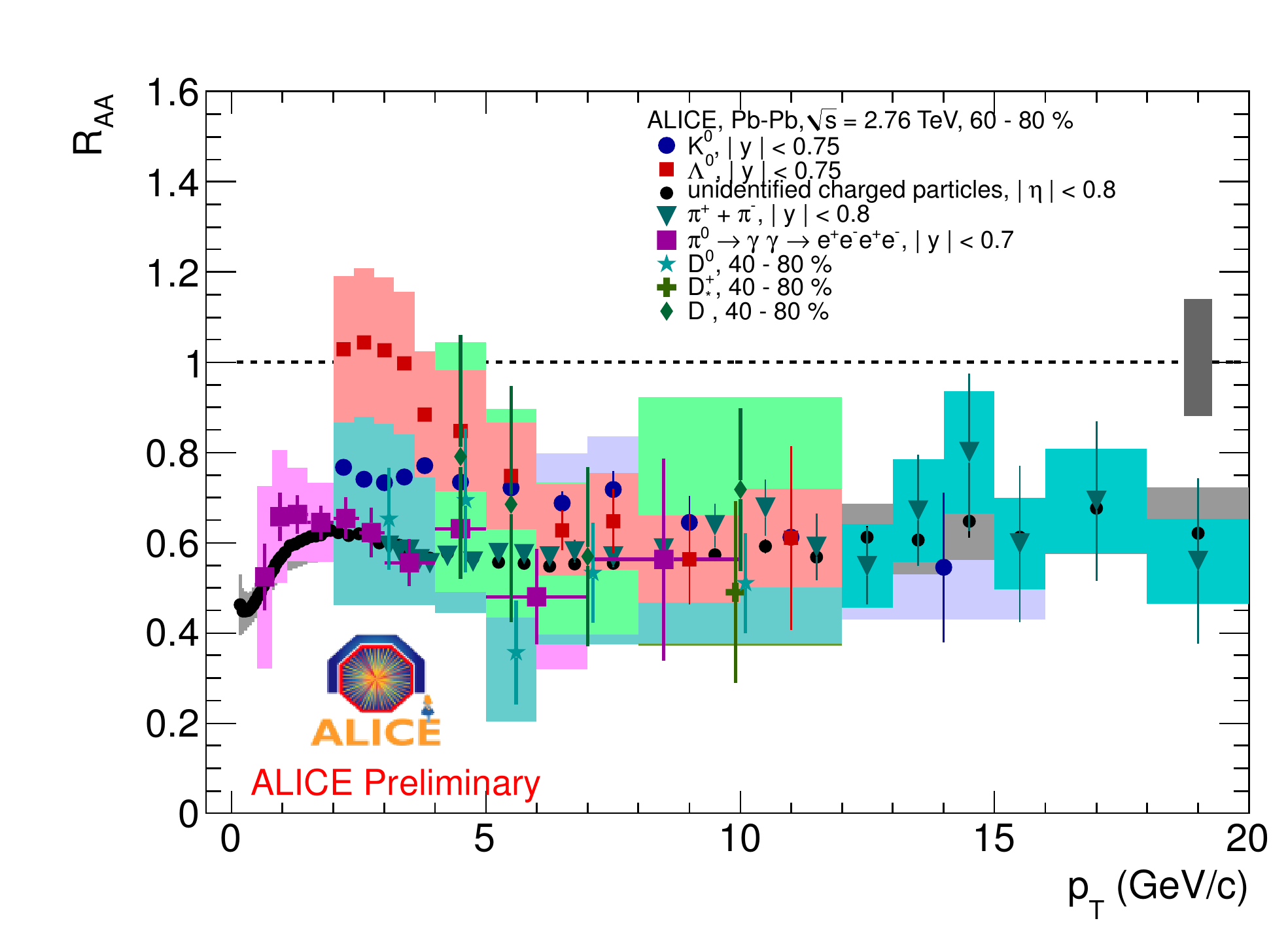}
  \caption{Nuclear modification factor $R_{AA}$ of charged particles,
    $K^0$, $\Lambda$, $\pi^\pm$, $D^+$, $D^0$, $D^{*+}$ in central
      (left) and peripheral (right) collisions.}
  \label{fig:PbPb-RAA_all}
\end{figure}

A nuclear modification factor $R_{AA}$ of charged pion production in
mid-rapidity (Fig.\ref{fig:PbPb-RAA_pions}) has lower values in the
range of moderate transverse momenta ($3<\pT<7-10$~GeV/$c$) than that
of unidentified charged particles, but at higher $\pT$ it coincides with
all charged particles.
To the contrary to charged pions, strange hadrons ($K^0_S$, $\Lambda$)
are less  suppressed in  the most central  collisions compared  to all
charged  particles (Fig.\ref{fig:PbPb-RAA_all}).  This is
explained by the  fact that strange quark production  is enhanced in a
hot  nuclear  medium,   and  this  strangeness  enhancement  partially
compensates energy loss of strange quarks, such that the overall
$R_{AA}$ value becomes larger than for pions. Lambda hyperons have no
suppression at $\pT<3-4$~GeV/$c$, which is interpreted by an
additional baryon enhancement in central heavy-ion collisions.

ALICE has reported also the first measurements of $D$ meson
suppression \cite{bib:PbPb-Dmesons} in Pb-Pb collisions in two
centrality classes, $0-20\%$ and $40-80\%$, shown in
Fig.\ref{fig:PbPb-RAA_all}. It was shown that the $R_{AA}$ values for
$D^0$, $D^+$ and $D^{*+}$ are consistent with each other within the
statistical and systematical uncertainties. Although the statistics of
the ALICE run 2010 is marginal for $D$ meson measurement, the obtained
result shows a hint that the $D$ mesons are less suppressed than
charged pions.
%

\section{Conclusion}

The ALICE collaboration is running an extensive research program with
proton-proton collisions. The domain where ALICE is competitive with
other LHC experiments, covers event characterization and identified
particle spectra at low and medium transverse momenta. Practically all
measured spectra in pp collisions at $\sqrt{s}=7$~TeV show
statistically significant deviations from models which well described
lower-energy results. Therefore new experimental results from pp
collision allow to tune various phenomenological models and pQCD
calculations.

A plenty of experimental results produced by the ALICE collaboration
from the first Pb-Pb data gives the first insight on strongly
interacting nuclear matter at the highest achievable collision
energy. It is evident that the quark-gluon matter produced in heavy
ion collision at LHC qualitatively has properties similar to what was
observed at RHIC. The matter produced at LHC has about 3 times larger
energy density, twice larger volume of homogeneity and about 20\%
larger lifetime. Like at RHIC, the matter at LHC reveals the
properties on an almost perfect liquid. Particle suppression appeared
to be stronger at LHC than at RHIC which is also an evidence of denser
medium produced at LHC.  At the end of 2011, LHC has delivered 10
times more data with Pb-Pb collision at $\sqsNN=2.76$~TeV, which will
bring more precise results.


%
%
\end{document}